\documentclass[12pt]{article}
\usepackage{hyperref}
\hypersetup{
colorlinks=true,
linkcolor=black,
citecolor=black,
urlcolor=cyan,
filecolor=black,
pdfborder={0 0 0},
}

\usepackage{amssymb}
\usepackage{amsfonts}
\usepackage{mathrsfs}
\usepackage{latexsym}
\usepackage{amsmath}
\usepackage{graphics}
\usepackage{graphicx,scalerel}
\newcommand\wh[1]{\hstretch{2}{\hat{\hstretch{.5}{#1\mkern3mu}}}\mkern-3mu}
\def\J{\wh{J}}
\usepackage{sectsty}
\usepackage{setspace}
\usepackage{multirow}

\usepackage[title]{appendix}
\vspace{.5in}

\sectionfont{\normalsize\bf}
\subsectionfont{\rm\normalsize}

\def\be{\begin{equation}}
\def\ee{\end{equation}}
\def\bd{\left|\begin{matrix}}
\def\ed{\end{matrix}\right|}
\def\half{{\scriptstyle {1\over 2}}}

\usepackage{cite}

\def\half{\textstyle{1\over 2}}
\def\quar{\textstyle{1\over 4}}

\numberwithin{equation}{section}

\begin{document}

\thispagestyle{empty}
\begin{flushright}
\end{flushright}
\baselineskip=16pt
\vspace{.5in}
{\begin{center}
{\bf Properties of the Conformal Yangian in Scalar and Gauge Field Theories}
\end{center}}
\vskip 1.1cm
\begin{center}
{Nikolaos Dokmetzoglou and Louise Dolan}
\vskip5pt

\centerline{\em Department of Physics and Astronomy}
\centerline{\em University of North Carolina, Chapel Hill, NC 27599} 
\vskip5pt

\bigskip
\bigskip
\bigskip
\bigskip
\end{center}

\abstract{\noindent 
Properties of the $SO(2,n)$ Yangian acting on scalar and gauge
fields are presented.
This differential operator representation of the
infinite-dimensional extension of the conformal algebra $SO(2,n)$
is proved to satisfy the Serre relation
for arbitrary spacetime dimension $n$ for off-shell scalar theory,
but only on shell and for $n=4$ in the gauge theory.
The $SO(2,n)$ Yangian acts
simply on the off-shell kinematic invariants
$(k_I+k_{I+1}+\ldots)^2$, and it annihilates individual off-shell
scalar $\lambda\phi^3$ Feynman tree graphs for $n=6$
if the differential operator representation is extended by graph
dependent evaluation
terms. The $SO(2,4)$ Yangian level one
generators are shown to act in a compact way on pure Yang-Mills gluon
tree amplitudes.
The action of the Yangian on the scattering polynomials of a CHY
formalism is also described.
}

\bigskip

\vskip120pt email: nidokme@live.unc.edu, ldolan@physics.unc.edu
\setlength{\parindent}{0pt}
\setlength{\parskip}{6pt}

\setstretch{1.05}
\vfill\eject
\vskip50pt
\section{Introduction}
\label{Introduction}

In this paper we prove various properties of
the Yangian algebra of the conformal group for scalar and gauge field
theories. Our focus is on the Yangian extension of $SO(2,n)$
and how this infinite-dimensional algebra acts on the kinematic invariants
and on tree level scattering amplitudes for both $\lambda\phi^3$
theory and pure Yang-Mills.
The $\lambda\phi^3$ theory of a single scalar field and
the non-abelian gauge theory are closely associated in the
Cachazo-He-Yuan (CHY) formalism \cite{CHY0,CHY1,CHY2},
since they share the same scattering polynomials.
It is well known that the Yangian of the superconformal algebra
$PSU(2,2|4)$ is a symmetry of planar ${\cal N}=4$ super Yang-Mills theory
\cite {DNW0, DNW1, Drummond0, Drummond1}, 
but less is known about how
the infinite-dimensional algebras act in a more realistic theory,
or in particular, in non-supersymmetric Yang-Mills.
To this end, we present certain features of the $SO(2,n)$ Yangian
in non-supersymmetric field theories.

In sections \ref{CYD} and \ref{Proof}, 
we discuss the differential operator representation
in momentum space of the $SO(2,n)$ Yangian, and describe the Serre
condition which this representation must satisfy for a consistent
algebraic structure.

In section \ref{Proof}, we prove the Serre relation for both the scalar
and gauge field theory representations, denoting what restrictions
apply.

In section \ref{ScalarTree}, it is shown that the $SO(2,n)$ Yangian
acts in a simple way on the kinematic invariants.
The  $\lambda\phi^3$ theory is conformally invariant at tree level in
$n=6$ dimensions, so the level zero generators annihilate both
partial amplitudes and individual Feynman graphs.
But the higher level Yangian generators only annihilate
individual Feynman graphs, and only when the representation of
the Yangian generators is extended by so-called evaluation terms.
That is to say, the $SO(2,6)$ Yangian is not a symmetry of $n=6$
$\lambda\phi^3$ theory, but it has some structure which may be useful.

In section \ref{Poly}, we look to extend our analysis by
presenting the action of the Yangian generators on the
off-shell scattering polynomials \cite{DGOS}. This will provide the Yangian
transformations of the graphs, since the momenta of a
scalar tree graph in the CHY formalism 
occur only in the scattering polynomials, and the Yangian generators act
only on the momenta.

In section \ref{hatPG3},
the action of the $SO(2,4)$ Yangian on pure Yang-Mills $N$-point
tree amplitudes
for $N=3,4$ is shown to have a compact form, albeit non-zero.
These expressions can also be expressed as traces of Dirac matrices,
which are shown to originate from tree amplitudes with 2 fermions and
$N-2$ gluons. This reflects the invariance of pure Yang-Mills gluon tree
amplitudes under the $PSU(2,2|4)$ Yangian, and may lead to some
interpretation of the role of supersymmetry in pure non-supersymmetric
gauge theory.

\section{The $SO(2,n)$ Conformal Algebra Yangian and its Defining Relations}
\label{CYD}

The $SO(2,n)$ conformal generators $J^{AB}$
satisfy the Lie algebra commutation relations
\begin{align}
[J^{AB},J^{CD}] &= -\eta^{AC} J^{BD}  -\eta^{BD} J^{AC}
+ \eta^{AD} J^{BC}  +\eta^{BC} J^{AD},\cr
J^{AB}&=-J^{BA}, \qquad 0\le A,B\le n+1,\qquad
\eta^{AB} = {\rm diagonal}\;(1,-1,-1,-1,\ldots,-1,1),\cr
g^{\mu\nu} &= \eta^{\mu\nu} = {\rm diagonal}\;(1,-1,-1,-1,\ldots,-1),
\qquad 0\le \mu,\nu\le n-1
\label{cY}\end{align}
The defining relations for the $SO(2,n)$ Yangian Hopf algebra \cite{DRIN1,
DRIN2} are
given in terms of the level zero generators $J^{AB}$, and the
level one generators $\wh{J}^{AB}$,
\begin{align}
&[J^{AB}, J^{CD}]= f^{ABCD}_{\hskip30pt EF} J^{EF},\;\qquad
[J^{AB}, \J^{CD}]= f^{ABCD}_{\hskip30pt EF} \J^{EF}\label{definingone}\\
\cr&[\J^{AB}, [ \J^{CD}, J^{EF}]]
+ [\J^{CD}, [ \J^{EF}, J^{AB}]]
+ [\J^{EF}, [ \J^{AB}, J^{CD}]]\cr
&=  {1\over 24}\; f^{AB}_{\hskip12pt GHMN}  f^{CD}_{\hskip12pt IJOP}
f^{EF}_{\hskip12pt KLQR} f^{MNOPQR}\; \{ J^{GH}, J^{IJ}, J^{KL}\}
\label{SR}\end{align}
where the structure constants of (\ref{cY}) can be displayed as
\begin{align}
f^{ABCD}_{\hskip30pt EF}
& = \half (-\eta^{AC}\delta^B_E\delta^D_F
-\eta^{BD}\delta^A_E\delta^C_F + \eta^{AD}\delta^B_E\delta^C_F
+\eta^{BC}\delta^A_E\delta^D_F\cr
&\hskip25pt
+ \eta^{AC}\delta^B_F\delta^D_E
+\eta^{BD}\delta^A_F\delta^C_E - \eta^{AD}\delta^B_F\delta^C_E
-\eta^{BC}\delta^A_F\delta^D_E)
\label{structure}\end{align}
and the indices are raised and lowered with the metric,
$f^{AB}_{\hskip12pt GHMN} = f^{ABG'H'}_{\hskip30pt MN} \;\eta_{GG'}\eta_{HH'}$
and
$f^{MNOPQR} = f^{MNOP}_{\hskip28pt Q'R'} \;\eta^{QQ'}\eta^{RR'}$
etc. The symmetrized triple product is defined by
\begin{align}
 \{ J^{GH}, J^{IJ}, J^{KL}\}&= J^{GH} J^{IJ} J^{KL}
+ J^{IJ} J^{GH} J^{KL} + J^{KL} J^{GH} J^{IJ}\cr
&\hskip20pt + J^{KL} J^{IJ} J^{GH} + J^{GH} J^{KL} J^{IJ}
+ J^{IJ} J^{KL} J^{GH}
\end{align}
The co-product construction of the Yangian provides for
a multi-site representation of the generators,
\begin{align}
J^{AB} = \sum_{i=1}^N J^{AB}_i,\qquad
\J^{AB} = \half f^{AB}_{\hskip12pt CDEF}\sum_{1\le i<j\le N} J_i^{CD} J_j^{EF}
\label{muls}\end{align}
So on a single site the level one generators $\J^{AB}=0$.
The relations (\ref{definingone}) follow from
the level zero single site commutation relations.
\begin{align}
[J_i^{AB},J_j^{CD}] &= \delta_{ij}
\; \Big( \;-\eta^{AC} J_i^{BD}  -\eta^{BD} J_i^{AC}
+ \eta^{AD} J_i^{BC}  +\eta^{BC} J_i^{AD}\;\Big)
\end{align}
The Serre relation (\ref{SR}) is satisfied if it holds for a single site,
namely
\begin{align} 0=
f^{AB}_{\hskip12pt GHMN}  f^{CD}_{\hskip12pt IJOP}
f^{EF}_{\hskip12pt KLQR} f^{MNOPQR}\; \{ J^{GH}, J^{IJ}, J^{KL}\}
\label{SSSR}\end{align}
Only certain representations of the single site
level zero generators will satisfy (\ref{SSSR}), and thus lead to a
consistent set of commutation relations. Once (\ref{SSSR}) is proved,
the defining relation (\ref{SR}) follows from the co-product.
The infinite number of higher level
generators of the Yangian algebra can then be derived
from commutators of the level one generators \cite{Bernard, Loebbert}.

In preparation for the proof,  we can reexpress the triple product
times the conformal structure constants in (\ref{SSSR}) as
the sum of three cyclic terms
\begin{align}
&f^{AB}_{\hskip12pt GHMN}  f^{CD}_{\hskip12pt IJOP}
f^{EF}_{\hskip12pt KLQR} f^{MNOPQR}\; \{ J^{GH}, J^{IJ}, J^{KL}\}
\cr =&\big[ 4\; \eta_{WY}\cr
&\cdot\Big(
\eta^{BD} \big( \{J^{FA},J^{EW},J^{YC}\}
+ \{J^{EC},J^{AW},J^{YF}\} - \{J^{FC},J^{EW},J^{YA}\}
- \{J^{EA},J^{CW},J^{YF}\}\big)\cr
&
-\eta^{AD} \big( \{J^{FB},J^{EW},J^{YC}\}
+ \{J^{EC},J^{B W},J^{YF}\} - \{J^{FC},J^{EW},J^{YB}\}
- \{J^{EB},J^{CW},J^{YF}\}\big)\cr
&
-\eta^{BC} \big( \{J^{FA},J^{EW},J^{YD}\}
+ \{J^{ED},J^{AW},J^{YF}\} - \{J^{FD},J^{EW},J^{YA}\}
- \{J^{EA},J^{DW},J^{YF}\}\big)\cr
&
+\eta^{AC}\big(\{J^{FB},J^{EW},J^{YD}\}
+\{J^{ED},J^{BW},J^{YF}\}-\{J^{FD},J^{EW},J^{YB}\}
-\{J^{EB},J^{DW},J^{YF}\}\big)\hskip-3pt\Big)\hskip-3pt\Big]\cr
&+ (ABCDEF\rightarrow CDEFAB) + (ABCDEF\rightarrow EFABCD)
\label{RHSG2}\end{align}
Then rewrite (\ref{RHSG2}) in terms of anti-commutators as
\begin{align}
=4\Bigg[&\Bigg(\Big({\scriptstyle{\eta^{DB}\big(
3 \,  (J^{FA} \{J^{EW},J^{YC}\}
+ J^{EC} \{J^{AW},J^{YF}\} - J^{FC} \{J^{EW},J^{YA}\}
- J^{EA} \{J^{CW},J^{YF}\}\,)\eta_{WY}}}\cr
& \hskip40pt {\scriptstyle{
- (\delta^W_W -6) (\eta^{EC}J^{FA} + \eta^{FA} J^{EC}
- \eta^{EA}J^{FC} - \eta^{FC} J^{EA}\,)\,}}\big)
- (A\leftrightarrow B) \Big)
- (C\leftrightarrow D) \Bigg)\cr
&+ (ABCDEF\rightarrow CDEFAB) + (ABCDEF\rightarrow EFABCD)\Bigg]
\label{rewr}\end{align}\normalsize
The terms proportional to $(\delta^W_W -6)$
will make zero contribution
in (\ref{rewr}) when all the permutations are performed, so we drop them.
The identities (\ref{RHSG2}) and (\ref{rewr})
hold for any number of sites, but we are interested to prove they vanish
on one site, {\it i.e.} the Serre condition (\ref{SSSR}) becomes
\begin{align}
&\Bigg[\Bigg(\Big(\eta^{DB} M^{EFCA}
- (A\leftrightarrow B)\; \Big)\quad - (C\leftrightarrow D) \Bigg)
\cr&\hskip20pt
+ (ABCDEF\rightarrow CDEFAB) + (ABCDEF\rightarrow EFABCD)\Bigg]
\;  =0
\label{sswr}\end{align}
where the single site tensor $M^{EFCA}$ is defined as
\begin{align}
&M^{EFCA}\cr &= \big(J_1^{FA} \{J_1^{EW},J_1^{YC}\}
+ J_1^{EC} \{J_1^{AW},J_1^{YF}\} - J_1^{FC} \{J_1^{EW},J_1^{YA}\}
- J_1^{EA} \{J_1^{CW},J_1^{YF}\}\,\big)\; \eta_{WY}
\cr\label{Mtensor}\end{align}

\section{Proof of the Serre Relation for the Differential Operator
Representation}
\label{Proof}

We consider the Serre relation
for the $SO(2,n)$ Yangian,
in the differential operator representation in momentum space,
which is relevant for scalar and spin one gauge fields with conformal spin $d$
in $n$ spacetime dimensions.
In this section we will
prove that (\ref{sswr}) is satisfied for off-shell scalar fields
with arbitrary $n$ and $d$. For gauge fields, we prove the
Serre relation (\ref{sswr}) is only satisfied for $n=4$, $d=1$, and on shell.
All fields are massless.

The differential operator representation for scalar and gauge fields in
momentum space is \cite{FP,CS}
\begin{align}
P_i^\mu &= k_i^\mu, \quad
L_i^{\mu\nu}= k_i^\mu\partial_i^\nu - k_i^\nu\partial_i^\mu
+ \Sigma^{\mu\nu}_i,\quad D = d +k_i^\nu\partial_{i\nu},\quad
\cr K_i^\mu &= 
2d \partial_i^\mu + 2 k_i^\nu\partial_{i\nu}\partial_i^\mu 
- k_i^\mu\partial_i^\nu
\partial_{i\nu} - 2 \Sigma^{\mu\nu}_i\partial_{i\nu}
\label{rep}\end{align} for each site $i$, where
\begin{align}
J_i^{\mu\nu}&= L_i^{\mu\nu},\qquad J_i^{n\mu}=\half (P_i^\mu - K_i^\mu),
\qquad J_i^{n+1,\mu}=\half (P_i^\mu + K_i^\mu),\qquad  J_i^{n,n+1} =D_i
\label{repi}\end{align}
The representation depends on the momenta $k_i^\mu$, $1\le i\le N$,
$0\le \mu\le n-1$,
their derivatives $\partial_{i\mu}\equiv {\partial\over\partial k_i^\mu}$
and a free real parameter $d$. 
For scalar fields $\Sigma^{\mu\nu} =0$.
For gauge fields $A_i^{\gamma_i},$ then $\Sigma^{\mu\nu}_{\alpha\gamma}
= (\delta^\mu_{\alpha} \delta^\nu_{\gamma} -
\delta^\nu_{\alpha} \delta^\mu_{\gamma})$.
So $L^{\mu\nu}_{i\;\alpha_i\gamma_i} A_i^{\gamma_i}
= (k_i^\mu\partial_i^\nu - k_i^\nu\partial_i^\mu) A_{i\alpha_i}
+  (\delta^\mu_{\alpha_i} \delta^\nu_{\gamma_i} -
\delta^\nu_{\alpha_i} \delta^\mu_{\gamma_i}) A_i^{\gamma_i}.$
See also Appendix \ref{Lev1}.

To prove the Serre relation for the representation (\ref{rep}),
first we write the level zero generators $J^{AB}$ and
their anticommutator $S^{AD}
\equiv \{J^{AB}, J^{CD}\}\eta_{BC}$ in terms of a smaller set of
operators, $\kappa^A,V^A$ and $\Sigma^{AB}$,
\begin{align}
J^{AB}&= \kappa^A V^B - \kappa^B V^A + \Sigma^{AB}\cr
S^{AD}&= -2\kappa^B\kappa_B  V^A V^D -2\; d\; \eta^{AD}
-2 \; (d-{\textstyle{n-2\over 2}})
(\kappa^A V^D + \kappa^D V^A)\cr
&\hskip12pt - 2\kappa^B V^A\Sigma^{CD}\eta_{BC}
- 2\kappa^B V^D\Sigma^{CA}\eta_{BC}
+ \Sigma^{AB}\Sigma^{CD}\eta_{BC}
+ \Sigma^{CD}\Sigma^{AB}\eta_{BC}
\label{JS}\end{align}
where
\begin{align}
\kappa^A&= (\kappa^\mu,\kappa^n,\kappa^{n+1}),
\qquad \kappa^\mu = P^\mu = k^\mu, \quad
\kappa^n = -(d+k^\rho\partial_\rho) = - \kappa^{n+1}\cr
V^A&= (V^\mu, V^n, V^{n+1}),\qquad V^\mu=\partial^\mu,\qquad
V^n= -\half (1 + \partial^\rho\partial_\rho),\quad
V^{n+1}= -\half (1 - \partial^\rho\partial_\rho)\cr
\Sigma^{AB}&= \delta^A_\mu\delta^B_\nu\Sigma^{\mu\nu}
+ \delta^A_\mu\;(\eta^{Bn}+\eta^{B,n+1})\;\Sigma^{\mu\rho}V_\rho
- \delta^B_\mu\;(\eta^{An}+\eta^{A,n+1})\;\Sigma^{\mu\rho}V_\rho\label{SIG}
\end{align}
satisfy a simpler algebra
\begin{align}
[\kappa^A, \kappa^B]&= \tilde c^{AB}_{\hskip10pt D} \;\kappa^D,\qquad
[V^A,V^B]=0,\qquad [\kappa^A, V^B] = -\eta^{AB} + c^{AB}_{\hskip10pt D}\;
V^D,\cr
\tilde c^{AB}_{\hskip10pt D} &= -\delta^A_D (\eta^{Bn}+\eta^{B,n+1})\;
+ \; \delta^B_D (\eta^{An}+\eta^{A,n+1})
=  -\tilde c^{BA}_{\hskip10pt D}
\cr
c^{AB}_{\hskip10pt D}&= -\delta^A_D (\eta^{Bn}+\eta^{B,n+1})\;
- \; \delta^B_D (\eta^{An}+\eta^{A,n+1}) =  c^{BA}_{\hskip10pt D}\cr
[\Sigma^{\mu\nu}, \Sigma^{\rho\sigma}] &=-\eta^{\mu\rho} \Sigma^{\nu\sigma}
-\eta^{\nu\sigma} \Sigma^{\mu\rho} +\eta^{\mu\sigma} \Sigma^{\nu\rho}
+\eta^{\nu\rho} \Sigma^{\mu\sigma},\qquad
[\Sigma^{\mu\nu}, \kappa^A] = [\Sigma^{\mu\nu}, V^A] =0\cr
\label{singelton}\end{align}
To prove ({\ref{sswr}) we consider the scalar and gauge cases separately.
For the scalar field, $\Sigma^{AB} = \Sigma^{\mu\nu} = 0,$
so (\ref{JS}) becomes
\begin{align}J^{AB}= \kappa^A V^B - \kappa^B V^A, \;
S^{AD}= -2\kappa^B\kappa_B  V^A V^D -2\; d\; \eta^{AD}
-2 \; (d-{\textstyle{n-2\over 2}})
(\kappa^A V^D + \kappa^D V^A)\label{scalJS}\end{align}
where for the remainder of this section all operators are at a single site,
but we suppress the single site notation.
To construct
$M^{EFCA}$, first compute
\begin{align}
-J^{EA} (\kappa^C V^F + \kappa^F V^C)
&= -(\kappa^E V^A- \kappa^A V^E)\; (\kappa^C V^F + \kappa^F V^C)\cr
&= -\kappa^E \kappa^C V^A V^F - \kappa^E \kappa^F V^A V^C
+ \kappa^A \kappa^C V^E V^F + \kappa^A \kappa^F V^E V^C\cr
&\hskip7pt - \kappa^E [V^A,\kappa^C] V^F
- \kappa^E [V^A,\kappa^F] V^C
\cr&\hskip7pt
+ \kappa^A [V^E,\kappa^C] V^F + \kappa^A [V^E,\kappa^F] V^C
\end{align} and combining as in the four tensor
\begin{align}
&\big( -J^{EA} (\kappa^C V^F + \kappa^F V^C) \; - A\leftrightarrow C\big)
\; - \; E\leftrightarrow F\cr
=&  - [\kappa^E, \kappa^C] V^A  V^F +  [\kappa^A, \kappa^F] V^E  V^C
+ [\kappa^E, \kappa^A] V^C  V^F -  [\kappa^C, \kappa^F] V^E V^A
\cr
&- J^{EC} [V^A,\kappa^F]
+ J^{AF} [V^E,\kappa^C]
+ J^{EA}  [V^C,\kappa^F]
- J^{CF}  [V^E,\kappa^A]\cr
=&- J^{EC} \eta^{AF}
+ J^{AF} \eta^{EC}
+ J^{EA} \eta^{CF}
-J^{CF} \eta^{EA}\cr
& +
(\eta^{En} + \eta^{E,n+1})\; 2 J^{AC}\; V^F
-  (\eta^{Fn} + \eta^{F,n+1})\; 2 J^{AC}\; V^E\cr
&-  (\eta^{Cn} + \eta^{,n+1})\; 2 J^{FE}\; V^A
+  (\eta^{An} + \eta^{A,n+1})\; 2 J^{FE}\; V^C
\label{s1}\end{align}
From the first term in $S^{AD}$, we have $\kappa^2 = \kappa^B\kappa_B
= k^2$,
\begin{align}
-J^{EA}\;\kappa^2 V^C V^F&=
-\kappa^2 J^{EA} V^C V^F -[J^{EA},\kappa^2] V^C V^F \cr
&= -\kappa^2 J^{EA} V^C V^F
 -(\eta^{En}+\eta^{E,n+1})\;\big( 2\kappa^2 V^A
+ 2(d-{\textstyle{n-2\over 2}})\;\kappa^A\big)\; V^C V^F
\cr &+  (\eta^{An}+\eta^{A,n+1})\;\big( 2\kappa^2 V^E
+ 2(d-{\textstyle{n-2\over 2}})\;\kappa^E\big)\; V^C V^F
\end{align}
with the permutations\begin{align}
&\Big(-J^{EA}\;\kappa^2 V^C V^F - A\leftrightarrow C\Big)
- E\leftrightarrow F
= \Big(-[J^{EA},\kappa^2] V^C V^F - A\leftrightarrow C\Big)
- E\leftrightarrow F\cr
&= -2\; (d- {\textstyle{\delta^\rho_\rho-2\over 2}})\;\;
\Big( (\eta^{En} + \eta^{E,n+1})\;  J^{AC}\; V^F
-  (\eta^{Fn} + \eta^{F,n+1})\;  J^{AC}\; V^E
\cr &\hskip100pt -  (\eta^{Cn} + \eta^{,n+1})\;  J^{FE}\; V^A
+  (\eta^{An} + \eta^{A,n+1})\;  J^{FE}\; V^C\Big)
\label{s2}\end{align}
Then combining (\ref{scalJS}), (\ref{s1}), (\ref{s2}),
we find the tensor
\begin{align}
M^{EFCA}&\equiv
-J^{EA} S^{CF} - J^{FC} S^{EA} +  J^{EC} S^{AF} +  J^{FA} S^{EC}\cr
&= (n-2) \; ( J^{EA}\eta^{CF} + J^{FC}\eta^{EA}
-J^{EC}\eta^{AF} - J^{FA}\eta^{EC}) 
\end{align}
Substituting this form for $M^{EFCA}$ into (\ref{sswr}) gives a
vanishing result due to the sum over various permutations. So
we have proved the Serre relation for scalar fields for
arbitrary $k^2$, $d$ and $n$, {\it i.e.} off-shell, for any conformal
dimension $d$ and spacetime dimension $n$.

\vskip10pt

For gauge fields, the proof of (\ref{sswr}) is more complicated.
The anti-commutator $S^{AD}$ has additional terms,  from (\ref{JS}),
which can be simplified using the commutation relations (\ref{singelton}),
\begin{align}
&- 2 \kappa^B V^A\Sigma^{CD}\eta_{BC}
- 2 \kappa^B V^D \Sigma^{CA}\eta_{BC}
+ \Sigma^{AB}\Sigma^{CD}\eta_{BC}
+ \Sigma^{CD}\Sigma^{AB}\eta_{BC}\cr
&=  \left( - 2 V^A\;k_\mu\;
\big (\delta^D_\rho + (\delta^D_{n+1}- \delta^D_n) V_\rho\big) \;
\Sigma^{\mu\rho} + \Sigma^{A\mu}\Sigma^{\nu D}\eta_{\mu\nu}\right)
+ A\leftrightarrow D
\label{ET}\end{align}
With the $\alpha,\gamma$ subscripts on $\Sigma^{\mu\nu}_{\alpha\gamma}$
displayed
explicitly,
(\ref{ET}) becomes
\begin{align}
&= \Big( - 2 k_{\alpha} \;V^A
\big (\delta^D_\gamma + (\delta^D_{n+1}- \delta^D_n) V_\gamma\big)
\cr
&\hskip20pt +(n-4)\big( \delta_\alpha^A
+ (\eta^{An} + \eta^{A, n+1})V_\alpha\big)
\;  \big( \delta^D_\gamma + (\delta^D_{n+1}- \delta^D_n)
V_\gamma\big)\cr
&\hskip20pt
- 2 V^A\; (\delta^D_{n+1}- \delta^D_n)  \eta_{\alpha\gamma}
+ 2V^A (\delta^D_\alpha + (\delta^D_{n+1}- \delta^D_n) V_\alpha\big)\; k_\gamma
\cr&\hskip20pt
+ (\delta^A_\delta + (\delta^A_{n+1}-\delta^A_n)V_\delta)\;
(\delta^D_{\delta'} + (\delta^D_{n+1}-\delta^D_n)V_{\delta'})\;
\eta^{\delta\delta'}\eta_{\alpha\gamma}\Big)
 + A\leftrightarrow D
\label{ET1}\end{align}
where we used the anti-commutator
\begin{align}
\Sigma^{A\mu}\Sigma^{\nu D}\eta_{\mu\nu} + A\leftrightarrow D&=
2 \; (\delta^A_\delta + (\delta^A_{n+1}-\delta^A_n)V_\delta)\;
(\delta^D_{\delta'} + (\delta^D_{n+1}-\delta^D_n)V_{\delta'})\;
\eta^{\delta\delta'}\eta_{\alpha\gamma}\cr
&\hskip7pt+(n-2)\;\Big(
\big(\delta^A_\alpha+(\delta^A_{n+1}-\delta^A_n)V_\alpha\big)
\;\big(\delta^D_{\gamma} + (\delta^D_{n+1}-\delta^D_n)V_{\gamma})\big)\cr
&\hskip70pt + \big(\delta^D_\alpha+(\delta^D_{n+1}-\delta^D_n)V_\alpha\big)
\;\big(\delta^A_{\gamma} + (\delta^A_{n+1}-\delta^A_n)V_{\gamma})\big)\Big)
\cr\end{align}
We reduce (\ref{ET1}) further using
\begin{align}
&V^A \; (\delta^D_{n+1}- \delta^D_n)   + V^D\; (\delta^A_{n+1}- \delta^A_n)
- (\delta^A_\delta + (\delta^A_{n+1}-\delta^a_n)V_\delta)\;
(\delta^D_{\delta'} + (\delta^D_{n+1}-\delta^D_n)V_{\delta'})\;
\eta^{\delta\delta'}\cr &= -\delta^A_{n+1}\delta^D_{n+1}
+ \delta^A_n\delta^D_n - \delta^A_\delta\delta^D_{\delta'}\eta^{\delta\delta'}
= -\eta^{AD}\end{align}
and combine it with the $\Sigma$- independent terms in (\ref{JS}) to evaluate
the anti-commutator for the gauge field representation as
\begin{align}
S^{AD}
&=\Big( - 2 k_\alpha \;V^A
\big (\delta^D_\gamma + (\delta^D_{n+1}- \delta^D_n) V_\gamma\big)
\cr
&\hskip15pt
+(n-4)\big( \delta_\alpha^A
+ (\eta^{An} + \eta^{A, n+1})V_\alpha\big)
\;  \big( \delta^D_\gamma + (\delta^D_{n+1}- \delta^D_n)
V_\gamma\big)\cr
&\hskip15pt
+ 2V^A (\delta^D_\alpha + (\delta^D_{n+1}- \delta^D_n) V_\alpha\big)\; k_\gamma
\cr
&\hskip15pt
+ \eta_{\alpha\gamma} \big(- k^2 V^A V^D  -\;(d-1)\; \eta^{AD}
-2 \; (d-{\textstyle{n-2\over 2}})\;\kappa^A V^D \big)\Big)
\cr & \hskip7pt + A\leftrightarrow D\cr
\label{Sgen}\end{align}
To construct the four tensor $M^{EFCA}$,
we first compute from (\ref{Sgen}) the product,
\begin{align}
-J^{EA} S^{CF} &= -J_{\alpha\beta'}^{EA}\;S^{CF}_{\beta\gamma}\eta^{\beta\beta'}
\cr
&= \Big( 2 J^{EA}_{\alpha\beta}\; k^\beta V^C
\big (\delta^F_\gamma + (\eta^{Fn} +\delta^{F,n+1}) V_\gamma\big)
+  J^{EA}_{\alpha\gamma}\;k^2\; V^C V^F\cr
&\hskip7pt - (n-4) J^{EA}_{\alpha\beta}\;
\big( \eta^{C\beta}
+ (\eta^{Cn} + \eta^{C, n+1})V^\beta\big)
\;  \big( \delta^F_\gamma + (\eta^{Fn}+ \eta^{Fn})
V_\gamma\big)\cr
&\hskip7pt -2  J^{EA}_{\alpha\beta}\;V^C \;
(\eta^{F\beta} +(\eta^{Fn}+\eta^{F,n+1})V^\beta)\; k_\gamma\cr
&\hskip7pt + (d-1) J^{EA}_{\alpha\gamma}\;\eta^{CF}
+ 2 (d-{\textstyle{n-2\over 2}})\; J^{EA}_{\alpha\gamma}\;
\kappa^C V^F\Big) \; + C\leftrightarrow F
\label{AT}\end{align}
Finally, with the use of
\begin{align}
J^{EA}_{\alpha\beta}\; k^\beta
&= [J^{EA}_{\alpha\beta}, k^\beta ] + k^\beta \;J^{EA}_{\alpha\beta}\cr
&= k_\alpha \;\Big( (\kappa^E + \eta^{En}+\eta^{E,n+1})\: V^A -
(\kappa^A + \eta^{An}+\eta^{A,n+1})\: V^E\Big)\cr
&\hskip7pt + (d+1-n)\; \Big( \delta^A_\alpha\;(\eta^{En}+\eta^{E,n+1})
-  \delta^E_\alpha\;(\eta^{An}+\eta^{A,n+1})\Big)\cr\cr
J^{EA}_{\alpha\gamma}\; k^2&=
[J^{EA}_{\alpha\gamma}, k^2] + k^2 \;J^{EA}_{\alpha\gamma}\cr
&= 2 (\eta^{En}+ \eta^{E,n+1})\;
(k_\alpha \delta^A_\gamma - \delta^A_\alpha k_\gamma)
- 2 (\eta^{An}+ \eta^{A,n+1})\;
(k_\alpha \delta^E_\gamma - \delta^E_\alpha k_\gamma)\cr
&\hskip7pt
+ \eta_{\alpha\gamma}\;
\Big( (\eta^{En}+ \eta^{E,n+1})\; ( 2 k^2 V^A + 2 (d-{\textstyle{n-2\over 2}})
\kappa^A) \cr
&\hskip50pt - (\eta^{An}+ \eta^{A,n+1})\; ( 2 k^2 V^E
+ 2 (d-{\textstyle{n-2\over 2}}) \kappa^E) \;\Big)\cr
&\hskip7pt +  k^2 \; J^{EA}_{\alpha\gamma}\cr\cr
J^{EA}_{\alpha\gamma}\; k^2\; V^C V^F
&= k_\alpha\; 2\;\big( (\delta^A_\gamma\; (\eta^{En}+ \eta^{E,n+1})-
\delta^E_\gamma\; (\eta^{An}+ \eta^{A,n+1})\big)\; V^CV^F\cr
&\hskip7pt - 2 \;\big( (\delta^A_\alpha\; (\eta^{En}+ \eta^{E,n+1})-
\delta^E_\alpha\; (\eta^{An}+ \eta^{A,n+1})\big)\;V^CV^F\; k_\gamma\cr
&\hskip7pt  + 2 \;\big( (\delta^A_\alpha\; (\eta^{En}+ \eta^{E,n+1})-
\delta^E_\alpha\; (\eta^{An}+ \eta^{A,n+1})\big)\cr
&\hskip20pt\cdot(\delta^C_\gamma + (\eta^{Cn}+\eta^{C,n+1})V_\gamma)\; V^F
+ V^C\; (\delta^F_\gamma + (\eta^{Fn}+\eta^{F,n+1})V_\gamma)\cr
&\hskip7pt
+ \Big[ \eta_{\alpha\gamma}\;
\Big( (\eta^{En}+ \eta^{E,n+1})\; ( 2 k^2 V^A + 2 (d-{\textstyle{n-2\over 2}})
\kappa^A) \cr &
\hskip50pt - (\eta^{An}+ \eta^{A,n+1})\; ( 2 k^2 V^E
+ 2 (d-{\textstyle{n-2\over 2}}) \kappa^E)\Big)\; \cr
&\hskip20pt +  k^2 \; J^{EA}_{\alpha\gamma}\Big] \; V^CV^F
\end{align}
the product becomes
\begin{align}
&-J^{EA} S^{CF} \cr &\hskip-20pt =
 \Big[ k_\alpha\; 2\; \big( (\kappa^E + \eta^{En}+\eta^{E,n+1})\: V^A -
(\kappa^A + \eta^{An}+\eta^{A,n+1})\: V^E\big)\cr
& + (2 (d+1-n) + 4)\; \big( \delta^A_\alpha\;(\eta^{En}+\eta^{E,n+1})
-  \delta^E_\alpha\;(\eta^{An}+\eta^{A,n+1})\big)\Big]\cr
&\hskip40pt \cdot
(\delta^C_\gamma + (\eta^{Cn}+\eta^{C,n+1})V_\gamma)\; V^F
+ V^C\; (\delta^F_\gamma + (\eta^{Fn}+\eta^{F,n+1})V_\gamma)\cr
& + k_\alpha\; 4\;\big( (\delta^A_\gamma\; (\eta^{En}+ \eta^{E,n+1})-
\delta^E_\gamma\; (\eta^{An}+ \eta^{A,n+1})\big)\; V^CV^F\cr
&- 4 \;\big( (\delta^A_\alpha\; (\eta^{En}+ \eta^{E,n+1})-
\delta^E_\alpha\; (\eta^{An}+ \eta^{A,n+1})\big)\;V^CV^F\; k_\gamma\cr
&
+ 2 \Big[{\tiny{ \eta_{\alpha\gamma}\;
\Big( (\eta^{En}+ \eta^{E,n+1})( 2 k^2 V^A + 2 (d-{\textstyle{n-2\over 2}})
\kappa^A)}}\cr&\hskip50pt {\tiny{ - (\eta^{An}+ \eta^{A,n+1})( 2 k^2 V^E
+ 2 (d-{\textstyle{n-2\over 2}}) \kappa^E)\Big)
+  k^2 J^{EA}_{\alpha\gamma}}} \Big] V^CV^F\cr
&+\Big[ \Big(- (n-4) J^{EA}_{\alpha\beta}\;
\big( \eta^{C\beta}
+ (\eta^{Cn} + \eta^{C, n+1})V^\beta\big)
\;  \big( \delta^F_\gamma + (\eta^{Fn}+ \eta^{Fn})
V_\gamma\big)
\cr &\hskip25pt
-2  J^{EA}_{\alpha\beta}\;V^C \;
(\eta^{F\beta} +(\eta^{Fn}+\eta^{F,n+1})V^\beta)\; k_\gamma
+ (d-1) J^{EA}_{\alpha\gamma}\;\eta^{CF}
+ 2 (d-{\textstyle{n-2\over 2}})\; J^{EA}_{\alpha\gamma}\;
\kappa^C V^F\cr
&\hskip20pt 
\; + C\leftrightarrow F\Big] \cr
\label{ntegen}\end{align}\normalsize
We see that (\ref{ntegen}) does not lead to a four tensor
(\ref{Mtensor}) that will satisfy the Serre relation for arbitrary $k^2, d, n$.
But for $k^2=0, d=1, n=4$, the product reduces to
\begin{align}
J^{EA} S^{CF}
&= 2 k_\alpha\;\Big[ (\kappa^E + \eta^{En}+\eta^{E,n+1})\; V^A -
(\kappa^A + \eta^{An}+\eta^{A,n+1})\; V^E\Big]\cr
&\hskip30pt \cdot
\Big[ (\delta^C_\gamma + (\eta^{Cn}+\eta^{C,n+1})V_\gamma)\; V^F
+ V^C\; (\delta^F_\gamma + (\eta^{Fn}+\eta^{F,n+1})V_\gamma)\Big]\cr
&\hskip7pt + 4 k_\alpha\; \big[ \delta^A_\gamma\; (\eta^{En}+ \eta^{E,n+1})-
\delta^E_\gamma\; (\eta^{An}+ \eta^{A,n+1}\big]\; V^CV^F\cr
&\hskip7pt - 4 \;\big[ \delta^A_\alpha\; (\eta^{En}+ \eta^{E,n+1})-
\delta^E_\alpha\; (\eta^{An}+ \eta^{A,n+1}\big]\;V^CV^F\;\; k_\gamma\cr
&\hskip7pt -2  J^{EA}_{\alpha\beta}\;\Big[ V^C \;
(\eta^{F\beta} +(\eta^{Fn}+\eta^{F,n+1})V^\beta)\;  + C\leftrightarrow F\Big]
\;k_\gamma
\label{JSf}\end{align}
which we recognize as a gauge transformation.

So $J^{EA} S^{CF} = 0$ when acting on on-shell gauge amplitudes
since they are gauge invariant. (\ref{JSf}) is a single site expression,
but it holds for any site. The gauge invariance of on-shell
amplitudes provides
$k_{i\gamma_i} A_N^{\gamma_1,\ldots,\gamma_N}
(k_1,\ldots, k_N)=0$ for any $i, 1\le i\le N$.
For example, if we consider (\ref{JSf}) at site one,
$k_{1\gamma} A_N^{\gamma,\alpha_2,\ldots,\alpha_N}(k_1,\ldots, k_N)=0$
will cause the terms in (\ref{JSf}) proportional to $k_\gamma$ to vanish.
The terms proportional to $k_\alpha$ will vanish upon multiplication
of $J^{EA}S^{CF}$ by $\epsilon^{\alpha}(k_1)$, since the
polarizations are transverse $k\cdot \epsilon(k) = 0$.
That is to say
\begin{align}\epsilon^\alpha (J^{EA}S^{CF})_{\alpha\gamma} A_N^{\gamma\ldots}
= 0
\end{align}
which implies $M^{EFCA}$ vanishes on the amplitudes, and proves
the Serre condition (\ref{sswr}) for the gauge field representation
(\ref{rep}) of the conformal Yangian on-shell when $n=4$, $d=1$.

This restriction to fields satisfying their free field equations of motion
for representation theory is familiar from earlier discussions of
the conformal group \cite{MT}.

Proof of the Serre relation in the context of the $SU(N)$ Yangian
and the $PSU(2,2|4)$ Yangian
was given in \cite{DNW1} using tensor operator methods.
These methods also occur in \cite{BLM}.
In this paper we emphasize that the $SO(2,n)$ Yangian gauge field
representation only has a consistent Serre relation for on-shell fields
and for $n=4$, $d=1$. In contrast, the scalar field
representation is consistent off shell,
for arbitrary $n$ and conformal dimension $d$.
\normalsize

\section{ Yangian Generators and the Scalar Tree Amplitudes}
\label{ScalarTree}

In this section, we compute how the $SO(2,n)$ Yangian generators
act on the kinematic invariants and the $\lambda\phi^3$
Feynman tree graphs.
In order to deal with the freedom in the amplitudes introduced by
momentum conservation, for arbitrary dimension $n$
we will consider the
amplitudes for the scalar theory
as functions of the ${N(N-3)\over 2}$ off-shell
variables $k_{[I,J]}^2$, for $1\le I< J<N$, but not $k^2_{[1,N-1]}$,
where the consecutive invariants are $k^2_{[I,J]}\equiv
(k_I + k_{I+1}+\ldots + k_J)^2$ as in \cite{DGOS}.

This set of invariants transforms in a simple way under the
level one generators $\hat P^\mu$,
that is to say merely with a multiplicative factor,
\begin{align}
-\hat P^\mu
&=\sum_{1\le i<j\le N} \;\big(P^\mu_i D_j + P_{i\rho}L^{\mu\rho}_j
- (i\leftrightarrow j)\big),\cr
-\hat P^\mu \; k^2_{[I,J]}
\cr &\hskip-50pt = \Big[ \Big( -2d\sum_{j=1}^N j k_j^\mu\Big) \;+
2 (k_1^\mu+k_2^\mu+\ldots+k_{I-1}^\mu
- k_{J+1}^\mu -k_{J+2}^\mu - \ldots -k_N^\mu)\Big] \; \; k^2_{[I,J]}
\label{prePs}\cr\end{align}
where we have assumed momentum conservation $\sum_{i=1}^N k_i^\mu =0$.
Apart from the common term proportional to $d$ that can be
extracted straightforwardly, we prove this as follows.

\begin{align}
&\sum_{1\le  i<j\le N} \;
( P_i^\mu \;k_j\cdot\partial_j
 + P_{i\rho} L_j^{\mu\rho} - (i\leftrightarrow j))
\; k^2_{[I,J]}\cr
&=\sum_{i=I}^J\Big[(k_1+\ldots + k_{i-1} - k_{i+1} - \ldots
- k_N)^\mu  k_i\cdot\partial_i \cr
&\hskip30pt
+ (k_1+\ldots + k_{i-1} - k_{i+1} - \ldots
- k_N)_\rho (k_i^\mu\partial_i^\rho - k_i^\rho\partial_i^\mu)\Big]
\;\; k^2_{[I,J]}\cr
&=2 \sum_{i=I}^J\Big[(k_1+\ldots + k_{i-1} - k_{i+1} - \ldots
- k_N)^\mu  k_i\cdot k_{[I,J]}\cr
&\hskip35pt
+ k_i^\mu (k_1+\ldots + k_{i-1} - k_{i+1} - \ldots
- k_N) \cdot k_{[I,J]} \cr
&\hskip35pt
- k_{[I,J]}^\mu (k_1+\ldots + k_{i-1} - k_{i+1} - \ldots
- k_N) \cdot  k_i\Big]
\label{Ps}\end{align}
We simplify each of the terms in (\ref{Ps}),
where the third term becomes
\begin{align}
&- 2\sum_{i=I}^J\; k_{[I,J]}^\mu (k_1+\ldots + k_{i-1} - k_{i+1} - \ldots
- k_N) \cdot  k_i\big)\cr
&= -2 k_{[I,J]}^\mu\;
(k_1+k_2+\ldots+k_{I-1}
- k_{J+1} -k_{J+2} - \ldots -k_N)\cdot k_{[I,J]}
\label{thirdT}\end{align}
and the first term contains the answer ({\ref{prePs})
plus a remainder,
\begin{align}
&2\sum_{i=I}^J\; (k_1+\ldots + k_{i-1} - k_{i+1} - \ldots
- k_N)^\mu  k_i\cdot k_{[I,J]} \cr
&= 2(k_1+k_2+\ldots+k_{I-1}
- k_{J+1} -k_{J+2} - \ldots - k_N)^\mu \; k^2_{[I,J]}\cr
&\hskip7pt +
2(-k_{I+1} - k_{I+2} - \ldots - k_J)^\mu\; k_I\cdot k_{[I,J]}
\; +\ldots + \;2 (k_I + \ldots \cr
&\hskip7pt + k_{J-1} -k_{J+1} - \ldots - k_J)^\mu\;
k_J\cdot k_{[I,J]}
\cr\label{firstT}\end{align}
Adding this remainder to  (\ref{thirdT}) and the second term in (\ref{Ps}),
we find the cancellation
\begin{align}
&
(-k_{I+1} - k_{I+2} - \ldots - k_J)^\mu\; k_I\cdot k_{[I,J]}
+ (k_I  -k_{I+2} - k_{I+3} - \ldots - k_J)^\mu\; k_{I+1}\cdot k_{[I,J]}
+ \cr
&\hskip-20pt {\scriptstyle{\ldots}} +
(k_I+ {\scriptstyle{\ldots}} + k_{J-1} -k_{J+1} - {\scriptstyle{\ldots}}
 - k_J)^\mu\; k_J\cdot k_{[I,J]}
+\hskip-5pt
\sum_{i=I}^J\;
k_i^\mu(k_1\hskip-3pt+
{\scriptstyle{\ldots}}\hskip-3pt + \hskip-3pt
k_{i-1} - k_{i+1} - {\scriptstyle{\ldots}}
- k_N)\hskip-3pt \cdot \hskip-3pt k_{[I,J]} \cr
&
-k_{[I,J]}^\mu\;
(k_1+k_2+\ldots+k_{I-1}
- k_{J+1} -k_{J+2} - \ldots -k_N)\cdot k_{[I,J]}\; =\; 0
\label{0T}\end{align}
\normalsize
by identifying the coefficient of each $k_i^\mu, I\le i\le J$
to be zero in (\ref{0T}). Here the consecutive momenta are
$k_{[I,J]}^\mu = (k_I+ k_{I+1}+\dots+k_J)^\mu.$
This proves (\ref{prePs}) which says
that $\hat P^\mu$ acts on off-shell invariants simply as multiplication
by a factor, somewhat similar to how the level zero generator
$P^\mu$ acts.

Since any $\lambda\phi^3$ Feynman tree graph is just the inverse
product of $(N-3)$ invariants, and the level one generator $\hat P^\mu$
is a first order differential operator, it is now easy to show how
$\hat P^\mu$ acts on the graphs.

We describe an off-shell  $\lambda\phi^3$ tree graph by
\begin{align}
\delta^n(k_1+\ldots +k_N) \;\; A_N^\Delta (k_1,\ldots, k_N)
\end{align}
where $\Delta$ denotes a subset of the off-shell invariants,
\begin{align}
A_N^\Delta (k_1,\ldots, k_N) = (-1)^{N+1}
\prod_{k_{[I,J]}^2\in\Delta} {1\over k^2_{[I,J]}}
\end{align}
for example for $N=4$, $\Delta$ is either $k^2_{[1,2]}$ or $k^2_{[2,3]}$.
For $N=5$, $\Delta$ is any of the five sets
$k^2_{[1,2]}, k^2_{[1,3]}$;
$k^2_{[1,2]}, k^2_{[3,4]}$;
$k^2_{[2,3]}, k^2_{[1,3]}$;
$k^2_{[2,3]}, k^2_{[2,4]}$;
$k^2_{[3,4]}, k^2_{[2,4]}$. These correspond to the two $N=4$ graphs
$-{1\over s_{12}}$, $-{1\over s_{23}}$ and the five $N=5$ graphs
${1\over s_{12}s_{123}},
{1\over s_{12}s_{34}},
{1\over s_{23} s_{123}},
{1\over s_{23}s_{234}},
{1\over s_{34}s_{234}},$ $s_{I,I+1,\dots, J} \equiv k^2_{[I,J]}$.
We can find the subsets for $N>5$ from
the off-shell recurrence relation \cite{DGOS},
\begin{align}
A_N^\Delta (k_1, \ldots, k_N)&= -{1\over s_{34}} A^{\Delta'}_{N-1}
(k_1,k_2, k_3+k_4, k_5,\ldots, k_N)
\label{recurr}\end{align}
where the momenta can be cycled to find all the $\Delta$.

The level one generator $\hat P^\mu$ acts as
\begin{align}
&-\hat P^\mu\; \;
\delta^n(k_1+\ldots +k_N) \; A_N^\Delta (k_1,\ldots, k_N)
\cr&=  \delta^n(k_1+\ldots +k_N) \cr
&\hskip20pt\cdot
\Big[ \Big( -2d\sum_{j=1}^N j\; k_j^\mu\Big) \;
- 2\sum_{[I,J]\atop k^2_{[I,J]}\in \Delta} (k_{[1, I-1]}^\mu
- k_{[J+1, N]}^\mu )\Big]\;\; A_N^\Delta (k_1,\ldots, k_N)
\cr
\end{align}
where $\hat P^\mu$ commutes through $\delta^n(k_1+\ldots+k_N)$ as in 
Appendix \ref{commutedelta}.
 
Clearly the level one generator does not annihilate the graph, and therefore
the conformal Yangian is not a symmetry of the $\lambda\phi^3$ scalar theory.
But the multiplicative factor is a sum of momenta $k_i^\mu$ with
various coefficients.
So we can define a set of evaluation parameters $c_{N,j}^\Delta$
for each individual graph,
\begin{align}
&c_{N,j}^\Delta= -2d \; j -2\sum_{[I,J]\atop k^2_{[I,J]}\in \Delta}  \Big\{
{1 \;\hbox{if $j\in [1, I-1]$}\atop -1  \;\hbox{if $j\in [J+1, N]$}}
\cr
&\hat P'^\mu \equiv \hat P^\mu + \sum_{j=1}^N c_{N,j}^\Delta \;P_j^\mu,\qquad
\hat P'^\mu\; \; \delta^n(k_1+\ldots +k_N) \; A_N^\Delta (k_1,\ldots, k_N)
=0
\label{Pprime}\end{align}
such that the level one generator shifted by the corresponding level
zero generator annihilates the graph.
Defining all shifted generators using the same parameters $c_{N,j}^\Delta$
\begin{align}
\J'^{AB}&\equiv \J^{AB} + \sum_{j=1}^N c_{N,j}^\Delta \;J^{AB}_j
\end{align}
it is straightforward to show that if
the $c_{N,j}^\Delta$ commute with $J^{AB}$, $\J^{AB}$
(which they do in this case because the parameters are constants), and if
$J^{AB}$, $\J^{AB}$ satisfy
the defining relations (\ref{definingone}), (\ref{SR}), then so do
$J^{AB}$, $\J'^{AB}$.
If we have further that all the level zero generators $J^{AB}$ annihilate
the graph, then (\ref{Pprime}) implies
\begin{align}
\J'^{AB} \; \; \delta^n(k_1+\ldots +k_N) \; A_N^\Delta (k_1,\ldots, k_N)
=0
\end{align}
This follows from the values of the conformal structure constants
(\ref{structure}).
See also \cite{Kaz},\cite{CKLMZ}.

In $\lambda\phi^3$ theory, $SO(2,n)$ is known to be a symmetry of the
classical Lagrangian in six dimensions, $n=6$,
where the field $\phi$ has conformal dimension $d=2$.
We can also check directly using the representation
(\ref{rep}), (\ref{repi}) that all tree graphs satisfy
\begin{align}
J^{AB} \; \; \delta^n(k_1+\ldots +k_N) \; A_N^\Delta (k_1,\ldots, k_N)
=0
\label{zerokill}\end{align}
For $P^\mu, L^{\mu\nu}$ this follows from momentum conservation,
and the Lorentz scalar nature of the kinematic invariants.
For the dilitation generator $D = \sum_{i=1}^N (d+ k_i\cdot \partial_i)$,
we see from the recurrence relation (\ref{recurr}) and the explicit forms for
$A^\Delta_4, A_5^\Delta$, that
$\sum_{i=1}^N k_i\cdot \partial_i  \;\;A_N^\Delta (k_1,\ldots, k_N) =
-2 (N-3)$, so
\begin{align}
&D\; \delta^n(k_1+\ldots +k_N) A_N^\Delta (k_1,\ldots, k_N)
\cr&= \Big( N(d-2) - n +6 \Big)\;\delta^n(k_1+\ldots +k_N)
A_N^\Delta (k_1,\ldots, k_N)
\label{DonA}\end{align}
where $D$ moves through the delta function as
$D\; \delta^n(k_1+\ldots +k_N) = \delta^n(k_1+\ldots +k_N) (-n + D)$.
\cite{GW} Clearly (\ref{DonA}) selects $n=6$, $d=2$ to kill the graphs.
Lastly the special conformal transformations
\begin{align}
K^\mu = \sum_{i=1}^N K_i^\mu =
\sum_{i=1}^N (2d \partial^\mu_i + 2 k_i\cdot\partial_i\partial_i^\mu
-k_i^\mu\partial^\rho_i\partial_{i\rho})
\end{align}
commute with the delta function as do $P^\mu, L^{\mu\nu}$.
The annihilation $K^\mu \;  A_N^\Delta (k_1,\ldots, k_N) =0$ results from the
recurrence relation as follows. If we can show
\begin{align}
&(K_3^\mu + K_4^\mu)\; (-{1\over s_{34}}) A_{N-1}^{\Delta'}
(k_1,k_2, k_3+k_4, k_5,\ldots, k_N)
\cr&= -{1\over s_{34}}\; K_{3+4}^\mu \;A_{N-1}^{\Delta'}
(k_1, k_2, k_3+k_4, k_5,\ldots, k_N)
\label{K34}\end{align}
then
\begin{align}&K^\mu \; (-{1\over s_{34}}) A_{N-1}^{\Delta'}
(k_1,k_2, k_3+k_4, k_5,\ldots, k_N)
\cr &= -{1\over s_{34}}
(K_1 + K_2 + K_{3+4} + K_5 + \ldots + K_N)^\mu A_{N-1}^{\Delta'}
(k_1, k_2, k_3+k_4, k_5,\ldots, k_N)\cr
\end{align}
which implies
\begin{align}
&K^\mu\;  A_N^\Delta (k_1,\ldots, k_N) \cr &=
-{1\over s_{34}}\;
(K_1 + K_2 + K_{3+4} + K_5 + \ldots +K_N)^\mu \;A_{N-1}^{\Delta'}
(k_1, k_2, k_3+k_4, k_5,\ldots, k_N) = 0
\cr\label{Krecurr}\end{align}
whenever $A_{N-1}^{\Delta'}$ is annihilated by its relevant
special conformal generators. This is explicitly true for $N=4,5$, so
the annihilation for higher $N$ follows iteratively.

To show (\ref{K34}), where
$K^\mu_{3+4} \equiv
(2d \partial^\mu_{3+4} + 2 (k_3+k_4)\cdot\partial_{3+4}\partial_{3+4}^\mu
-(k_3+k_4)^\mu\partial^\rho_{3+4}\partial_{3+4,\rho})$, we find,
using $\partial_{3+4}^\mu\equiv {\partial\over \partial(k_3+k_4)_\mu}$,
\begin{align}
&(K_3^\mu + K_4^\mu)\; (-{1\over s_{34}}) A_{N-1}^{\Delta'}
(k_1,k_2, k_3+k_4, k_5,\ldots, k_N)
\cr
&= -{1\over s_{34}}\Big[(4d - 4) \partial_{3+4}^\mu
-{(8d -2n -4) (k_3+k_4)^\mu \over s_{34}}
+ 2(k_3+k_4)\cdot \partial_{3+4} \partial_{3+4}^\mu
\cr &\hskip40pt 
- (k_3+k_4)^\mu \partial^\rho_{3+4} \partial_{\rho,3+4}\Big]
\; A_{N-1}^{\Delta'} (k_1,k_2, k_3+k_4, k_5,\ldots, k_N)
\end{align}
which gives (\ref{K34}) for $n=6,d=2$.

So any tree graph for $\lambda\phi^3$ theory
is annihilated by the $SO(2,6)$ Yangian where the representation of the
level one generators $\J'^{AB}$ depend on the graph.
We look next briefly at the CHY formalism for these graphs,
where the Yangian generators act solely through the scattering polynomials.
\normalsize

\section{Yangian Properties of the Scattering Polynomials}
\label{Poly}

In the CHY formalism \cite{CHY0, CHY1, CHY2},
off-shell tree level amplitudes can be described as contour integrals
encircling the zeros of the scattering polynomials \cite{DGOS}.
These polynomials are functions of the variables $z_1,\ldots, z_N$
with coefficients given by the set of off-shell invariants $k^2_{[I,J]}$,
$1\le I<J\le N$ discussed in the previous section.
The momentum dependence of $\lambda\phi^3$ amplitudes is entirely
via the polynomials. Thus the action of the Yangian generators
on the scalar amplitudes employs how the Yangian acts on
the scattering polynomials.

The off-shell scattering polynomials $h_m^N$, $1\le m\le N-3$ are
\cite{DGOS}
\begin{align}\label{Mfix}
h^N_m&=\sum_{J=2}^{N-2}k^2_{[1,J]}\, (z_{J}-z_{J+1})\Pi_{[1,J]^o}^{m-1}
-  \sum_{J=3}^{N-1}k^2_{[2,J]}\,(z_{J}-z_{J+1})\Pi_{[2,J]^o}^{m-1}\cr
&\qquad+ \sum_{3\leq I<J<N}k^2_{[I,J]}\,(z_{I}-z_{I-1}) (z_{J}-z_{J+1})
\Pi_{[I,J]^{\prime o}}^{m-2}\,,
\end{align}
$1\leq m\leq N-3$, where $[I,J]^{\prime o}=[I,J]^o\cap A',$
$A'= \{2,3,\ldots, N\}$	and $[I,J]^o$ is the complement of $\{I-1,I,J+1\}$
in $A\equiv\{1,2,\ldots,N\}$. $\Pi^n_U $ is the homogeneous symmetric
polynomial 
\be
\Pi_U^n=\sum_{i_1<i_2<\cdots<i_n\atop i_a\in U}z_{i_1}z_{i_2}\cdots z_{i_n}
\nonumber\ee
where $U\subset A$ and $n\leq |U|$,  $A\equiv \{1,2,\cdots,N\}$.

Examples for $N=4$, $N=5$ are
\begin{align}
h_1^4&= s_{12} - z_3 (s_{12} + s_{23}) \cr
h_1^5&= s_{12} + z_3 (s_{123} -s_{23} -s_{12}) + z_4 (s_{23} - s_{123}
- s_{234})\cr
h_2^5&= z_3 s_{123} + z_4 (s_{12} -s_{123} -s_{34}) + z_3z_4
(s_{34} - s_{12} -s_{234})\cr
\end{align}
The level zero generators $L^{\mu\nu}$, $D$ act simply,
\begin{align}
L^{\mu\nu} \; h_m^N =0,\qquad D \; h_m^N = (Nd +2)\; h_m^N
\end{align}
The off-shell $\lambda\phi^3$ partial amplitudes are given by
\begin{align}
A_N^{\rm partial}(k_1,\ldots,k_N)&=
\oint {1\over z_{N-1}}\prod_{m=1 }^{N-3}{1\over h^N_m
(z,k)}\prod_{2\leq a<b\leq N-1} (z_a-z_b)\prod_{a=2}^{N-2}{z_adz_{a+1}
\over (z_a-z_{a+1})^2}.\label{amp2}
\end{align}
and the individual Feynman graphs are given by similar formulae
in terms of cross ratios \cite{DGOS}.

All of these contain the momenta through inverse products of the
polynomials. So the invariance under $P^\mu, L^{\mu\nu},
D$ for $n=6$, $d=4$ follows from
\begin{align}
P^\mu\;\Bigg[ \delta^n(\Sigma_{j=1}^N k_j)\;\prod_{m=1}^{N-3} {1\over
h_m^N} \Bigg]=
L^{\mu\nu}\;\Bigg[ \delta^n(\Sigma_{j=1}^N k_j)\;\prod_{m=1}^{N-3} {1\over
h_m^N} \Bigg]=0\cr
D\;\Bigg[ \delta^n(\Sigma_{j=1}^N k_j)\;\prod_{m=1}^{N_3} {1\over
h_m^N} \Bigg]
= (N(d-2)+6-n)\;\;\delta^n(\Sigma_{j=1}^N k_j)\;\prod_{m=1}^{N_3} {1\over
h_m^N}
\end{align}
The special conformal generator $K^\mu$ is not first order in the derivative
operators. We will show how it acts on the contour integral for the
two $N=4$ Feynman graphs. Its action on higher $N$ involve
multivariable contour integrals, but they must also vanish for
$n=6$, $d=2$, in accordance with (\ref{Krecurr}). Let $h=h_1^4
= s_{12} - z(s_{12} + s_{23})$,
\;$z=z_3$,
\begin{align}
K^\mu\; {1\over h} &=
{(-8d +2n+4)\;\big[ (1-z) (k_1+k_2)^\mu - z (k_2+k_3)^\mu\big]\over
(h)^2}\cr
&\hskip10pt + {8 z (1-z)\;\big[ (k_3^2-k_2^2) k_1^\mu
+ (k_4^2-k_2^2) k_2^\mu + (k_1^2-k_2^2) k_3^\mu\big]\over (h)^3}
\label{Kh}\end{align}
The $N=4$ individual graphs are
\begin{align}
-{1\over s_{12}} &= \oint_{h=0} {dz\over h}\;{1\over z},
\qquad\quad  -{1\over s_{23}} = \oint_{h=0} {dz\over h}\;{1\over (1-z)}
\end{align}
From (\ref{Kh}) the surviving integrals are
\begin{align}
K^\mu\; \oint_{h=0} {dz\over h}\;{1\over z}&=
(-8d + 2n +4) (k_1+k_2)^\mu\oint_{h=0} {dz\over( h)^2}\;{1\over z}\cr
&=(8d - 2n -4) (k_1+k_2)^\mu\oint_{z=0} {dz\over z}\;{1\over (h)^2}
= {(8d - 2n -4) (k_1+k_2)^\mu\over s_{12}^2},\cr
K^\mu\; \oint_{h=0} {dz\over h}\;{1\over (1-z)}&=
(8d - 2n -4) (k_2+k_3)^\mu\oint_{h=0} {dz\over( h)^2}\;{z\over (1-z_3)}\cr
&\hskip-40pt
=(-8d + 2n +4) (k_2+k_3)^\mu\oint_{z=1} {dz\over (1-z)}\;{1\over (h)^2}
= {(8d - 2n -4) (k_2+k_3)^\mu\over s_{23}^2}\cr
\label{Khh}\end{align}
In the integrands with ${1\over (h)^3}$ the contour encircles all the
zeros of the denominator, so those integrals vanish.
In (\ref{Khh}) we can swap the contour around $h=0$ to minus the
contour around $z=0$, etc.  since there is no residue at infinity.

For $N=5$, acting with $K^\mu$ will result in terms like
$\oint_{h_1=h_2=0}{dz_3dz_4\over h^2_1 h_2 z_4}\equiv \hbox{Res}(h_1^2, h_2)$,
which can be evaluated
with the global residue theorem, by dividing the factors in the denominator
into two disjoint sets, $\{h_1^2, z_4\}$ and $\{ h_2\}$, so that
$\hbox{Res}(h_1^2, h_2) = - \hbox{Res}(z_4, h_2).$

The level one generators acting on the scattering polynomials
are similarly complicated.
We give the expression for
$\hat P^\mu$ on $h= h_1^4$,
\begin{align}
&-\hat P^\mu\; h = -2k_3^\mu s_{12} (1-z) - 2 k_1^\mu s_{23} z
+h \; (- 2k_4^\mu + d\; (3k_1 +  k_2 - k_3 -3k_4)^\mu )\cr
&-\hat P^\mu\;{1\over h} = {1\over h^2}\;
[2k_1^\mu s_{23} z + 2 k_3^\mu s_{12} (1-z)]\; + 
{1\over h}\; (2k_4^\mu  + d\; (3k_1 +  k_2 - k_3 -3k_4)^\mu )
\end{align}
The polynomial does not transform simply.

The partial amplitude transforms as expected,
\begin{align}
&-\hat P^\mu\;\oint {dz\over h} \; ({1\over z} + {1\over 1-z})
\cr
&=  2k_3^\mu s_{12}  \oint {dz\over h^2_1} {1\over z}
+ 2k_1^\mu s_{23} \oint {dz\over h^2_1} {1\over 1-z} + 2k^\mu_4
\oint {dz\over h_1} \; ({1\over z} + {1\over 1-z}) + {\cal O}(d)\cr
&= -{2 k_3^\mu \over s_{12}} + {2 k_1^\mu\over s_{23}}
+ 2k^\mu_4 \;(-{1\over s_{12}} - {1\over s_{23}}) + {\cal O}(d)
\label{4N}
\end{align}

For gauge theory, the CHY integral has also a Pfaffian
in the numerator and it depends on momenta. 
For conformally invariant
amplitudes, including the Pfaffian has the effect of dropping the
spacetime dimension from six to four. This reflects that the Yangian
generators must now act on both the polynomials and the Pfaffian.

\section{$SO(2,4)$ Yangian Level One Generator Acting on
Gluon Tree Amplitudes}
\label{hatPG3}

We have computed explicitly how the level one $SO(2,4)$ Yangian generator
$\hat P^\mu$ acts on gluon tree amplitudes in pure Yang-Mills theory,
using the momentum space
differential operator representation of $\hat P^\mu$. Although
the generator does not annihilate the gluon amplitudes,
we find a somewhat compact form.
We show this for
the 3-gluon amplitude and the 4-gluon partial amplitude, on shell.
This could be useful to understand
what role the $SO(2,4)$ Yangian might play in pure Yang-Mills.
Since it does not annihilate the amplitude, it is not a symmetry.
But it may serve some function as a spectrum generating algebra would.
We will show that our answer can also be written in terms of traces of
Dirac gamma matrices, which is motivated by the fact that the
supersymmetic $PSU(2,2|4)$ Yangian does annihilate the superamplitude.
In particular, the $PSU(2,2|4)$ Yangian level one generators
annihilate the pure gluon amplitudes, and
the $SO(2,4)$ Yangian level one generator acting on the
gluon amplitude changes it into an amplitude involving 2 fermions and
2 less gluons.
This may help us to interpret a possible role of
supersymmetry in non-supersymmetric gauge theory.
But this is beyond the scope of this paper. 

\vskip20pt
The three-gluon tree amplitude is
\begin{align}
A^{a\alpha_1,\,b\alpha_2, \,c\alpha_3}
&= - g\; f_{abc}\;
(2\pi)^4 \delta^4(k_1+k_2+k_3)\;\cr 
&\hskip7pt\cdot\Big(
\eta^{\alpha_2\alpha_3} (k_2-k_3)^{\alpha_1}
+ \eta^{\alpha_3\alpha_1} (k_3-k_1)^{\alpha_2}
+ \eta^{\alpha_1\alpha_2} (k_1-k_2)^{\alpha_3}\Big)
\cr\label{threeg}\end{align}

For $N=3$ the $SO(2,4)$ Yangian level one generator for the gauge theory is
\begin{align}
&-\hat P^{\mu\;\alpha_1\alpha_2\alpha_3}_{\gamma_1\gamma_2\gamma_3} =
(P_1^\mu (D_2+D_3) + P_2^\mu D_3 - (P_2+P_3)^\mu D_1 - P_3^\mu
D_2) \; \delta^{\alpha_1}_{\gamma_1} \delta^{\alpha_2}_{\gamma_2}
\delta^{\alpha_3}_{\gamma_3} \cr
&\hskip90pt + (P_{1\rho} (L_2+L_3)^{\mu\rho} + P_{2\rho} L_3^{\mu\rho}
- (P_2+P_3)_\rho L_1^{\mu\rho} -P_{3\rho} L_2^{\mu\rho})^{\alpha_1\alpha_2
\alpha_3}_{\gamma_1\gamma_2\gamma_3} \cr
&= \Big(2 (k_1-k_3)^\mu
-(k_2+k_3)^\mu k_1\cdot\partial_1
+ (k_1-k_3)^\mu k_2\cdot\partial_2
+ (k_1+k_2)^\mu k_3\cdot\partial_3\cr
&\hskip7pt -(k_2+k_3)_\beta (k_1^\mu\partial_1^\beta - k_1^\beta\partial_1^\mu)
+(k_1-k_3)_\beta (k_2^\mu\partial_2^\beta - k_2^\beta\partial_2^\mu)
\cr &\hskip7pt
+ (k_1+k_2)_\beta (k_3^\mu\partial_3^\beta - k_3^\beta\partial_3^\mu)
\Big) \;\;\;
\delta^{\alpha_1}_{\gamma_1} \delta^{\alpha_2}_{\gamma_2}
\delta^{\alpha_3}_{\gamma_3} \cr
&\hskip7pt  -(k_2+k_3)_\beta (\eta^{\mu\alpha_1}\delta^\beta_{\gamma_1}
- \eta^{\beta\alpha_1}\delta^\mu_{\gamma_1}) \; \delta^{\alpha_2}_{\gamma_2}
\delta^{\alpha_3}_{\gamma_3}
+\hskip-3pt 
(k_1-k_3)_\beta (\eta^{\mu\alpha_2}\delta^\beta_{\gamma_2}
- \eta^{\beta\alpha_2}\delta^\mu_{\gamma_2}) \; \delta^{\alpha_1}_{\gamma_1}
\delta^{\alpha_3}_{\gamma_3}\cr
&\hskip7pt + (k_1+k_2)_\beta (\eta^{\mu\alpha_3}\delta^\beta_{\gamma_3}
- \eta^{\beta\alpha_3}\delta^\mu_{\gamma_3}) \; \delta^{\alpha_1}_{\gamma_1}
\delta^{\alpha_2}_{\gamma_2}
\end{align}
This level one generator moves trivially
through the delta function, as shown in Appendix A, and we find
the compact formula
\begin{align}
&-
\hat P^{\mu\alpha_1\alpha_2\alpha_3}_{\gamma_1\gamma_2\gamma_3}
\; A^{a\gamma_1,\;b\gamma_2,\;c\gamma_3}
\cr
&=- g\;f_{abc}\; (2\pi)^4 \delta^4(k_1+k_2+k_3)\cr
&\cdot \Big( \; 4 k_1^\mu\;
\big( \eta^{\alpha_2\alpha_3} k_2^{\alpha_1}
+\eta^{\alpha_3\alpha_1} k_3^{\alpha_2}
+2\eta^{\alpha_1\alpha_2} k_1^{\alpha_3}\big)
- 4 k_3^\mu\;
\big( 2\eta^{\alpha_2\alpha_3} k_2^{\alpha_1}
+\eta^{\alpha_3\alpha_1} k_3^{\alpha_2}
+\eta^{\alpha_1\alpha_2} k_1^{\alpha_3}\big)
\cr &\hskip20pt + \eta^{\mu\alpha_1}\; (
-4 k_1^{\alpha_2} k_1^{\alpha_3})
+ \eta^{\mu\alpha_2}\; (-4 k_2^{\alpha_1} k_2^{\alpha_3})
+ \eta^{\mu\alpha_3}\; (-4 k_3^{\alpha_1} k_3^{\alpha_2})\;\Big)\cr
\label{PA3f}\end{align}
Since this expression is not zero, we know the $SO(2,4)$ Yangian
is not a symmetry of pure Yang-Mills theory. But it shows how
this Yangian acts in a non-supersymmetric gauge theory.
The expression is gauge invariant, {\it i.e.} it vanishes when multiplied
by any $k_i^{\alpha_i}$ for  $1\le i\le 3$.
It is not cyclic invariant since $\hat P^\mu$ is not cyclic.
We assume transverse polarizations $k_i\cdot \epsilon_i(k_i)=0,$
and drop terms in (\ref{PA3f}) proportional to
$k_i^{\alpha_i}$ since they correspond to gauge transformations.
The on-shell conditions $k_i^2=0$ in four spacetime dimensions
are required from the Serre relation.
The presence of the $\eta^{\mu\alpha_i}$ is necessary for gauge invariance;
and these fixed tensor
terms cannot be removed by extending $\hat P^\mu$ by evaluation
parameters, as they could in the case of scalar $\lambda\phi^3$ theory.

To pursue a simpler form of (\ref{PA3f}),
we note it can be streamlined from the equivalence
\begin{align}
&\delta^4(k_1+k_2+k_3)\cr
&\cdot \Big( \; 4 k_1^\mu\;
\big( \eta^{\alpha_2\alpha_3} k_2^{\alpha_1}
+\eta^{\alpha_3\alpha_1} k_3^{\alpha_2}
+2\eta^{\alpha_1\alpha_2} k_1^{\alpha_3}\big)
-  4 k_3^\mu\;
\big( 2\eta^{\alpha_2\alpha_3} k_2^{\alpha_1}
+\eta^{\alpha_3\alpha_1} k_3^{\alpha_2}
+\eta^{\alpha_1\alpha_2} k_1^{\alpha_3}\big)
\cr&\hskip20pt + \eta^{\mu\alpha_1}\; (
-4 k_1^{\alpha_2} k_1^{\alpha_3})
+ \eta^{\mu\alpha_2}\; (-4 k_2^{\alpha_1} k_2^{\alpha_3})
+ \eta^{\mu\alpha_3}\; (-4 k_3^{\alpha_1} k_3^{\alpha_2})\;\Big)\cr
&= \delta^4(k_1+k_2+k_3)\;
\big(tr (\gamma^{\alpha_2}\gamma^\zeta\gamma^{\alpha_3}
\gamma^\omega\gamma^{\alpha_1}\gamma^\mu) k_{1\omega}k_{2\zeta}
\cr &\hskip100pt
- tr(\gamma^{\alpha_3}\gamma^\zeta\gamma^{\alpha_2}
\gamma^\omega\gamma^{\alpha_1}\gamma^\mu)
k_{1\omega} k_{3\zeta}
+tr(\gamma^{\alpha_3}
\gamma^\zeta\gamma^{\alpha_1}\gamma^\omega\gamma^{\alpha_2}
\gamma^\mu)  k_{2\omega}k_{3\zeta}\big)
\label{pmub}\cr\end{align}
Here the Dirac $\gamma$ matrices are in a Weyl representation,
\begin{align}
\gamma^\mu&=\left(\begin{matrix}0&\sigma^\mu\cr
\bar\sigma^\mu&0\end{matrix}\right),\;\sigma^\mu = (1,\sigma^i),\;
\bar\sigma^\mu = (1,-\sigma^i),\cr
\{\gamma^\mu,\gamma^\nu\}
&= 2\eta^{\mu\nu},\;\eta^{\mu\nu} = {\rm diag}\;(1,-1,-1,-1)
\label{gm1}\end{align}
The equivalence can be checked using standard trace formulae,
\begin{align}
&tr (\gamma^\mu\gamma^\nu) = 4 \eta^{\mu\nu}, \qquad
tr (\gamma^\mu\gamma^\nu\gamma^\rho\gamma^\lambda)
= 4 \;\big(\eta^{\mu\nu}\eta^{\rho\lambda} -  \eta^{\mu\rho}\eta^{\nu\lambda}
+  \eta^{\mu\lambda}\eta^{\nu\rho}\big)\cr
&tr (\gamma^\mu\gamma^\nu\gamma^\rho\gamma^\lambda\gamma^\omega\gamma^\zeta)
=\eta^{\mu\nu} tr(\gamma^\rho\gamma^\lambda\gamma^\omega\gamma^\zeta)
- \eta^{\mu\rho} tr(\gamma^\nu\gamma^\lambda\gamma^\omega\gamma^\zeta)
+ \eta^{\mu\lambda} tr(\gamma^\nu\gamma^\rho\gamma^\omega\gamma^\zeta)
\cr &\hskip100pt
- \eta^{\mu\omega} tr(\gamma^\nu\gamma^\rho\gamma^\lambda\gamma^\zeta)
+ \eta^{\mu\zeta} tr(\gamma^\nu\gamma^\rho\gamma^\lambda\gamma^\omega)\cr
\label{stantr}\end{align}
which can be extended to the trace of products of any number of
$\gamma$ matrices using the anticommutator.
We were motivated to find the identity (\ref{pmub}) by extending the
level one generator to include supercharges.
In position space we would have
\begin{align}
&\langle 0|T A^{a\gamma_1}(x_1) A^{b\gamma_2}(x_2) A^{c\gamma_3}(x_3)|0\rangle
= G^{a\gamma_1, b\gamma_2, c\gamma_3}(x_1 x_2 x_3)
\end{align}
and for $PSU(2,2|4)$ Yangian invariance of the three-gluon tree amplitude
we assume a level one generator of the form
\begin{align}
&- \hat P^{\mu\alpha_1\alpha_2\alpha_3}_{x,SS\;\gamma_1\gamma_2\gamma_3}
\; G^{a\gamma_1, b\gamma_2, c\gamma_3}(x_1 x_2 x_3) = 0
\cr&=\hskip-5pt
\sum_{1\;\le i<j\le 3}\Big( P_i^\mu D_j + P_{i\rho}L_j^{\mu\rho}
-\quar \bar\sigma^{\mu\dot\alpha\alpha}
Q^A_{\alpha i} \tilde Q_{A\dot\alpha j} - (i\leftrightarrow j)
\Big )^{\alpha_1\alpha_2\alpha_3}_{x\;\gamma_1\gamma_2\gamma_3}
\hskip-20pt G^{a\gamma_1, b\gamma_2, c\gamma_3}(x_1 x_2 x_3)\cr
&=\sum_{1\;\le i<j\le 3}\Big( P_i^\mu D_j + P_{i\rho}L_j^{\mu\rho}
- (i\leftrightarrow j)
\Big)^{\alpha_1\alpha_2\alpha_3}_{x\;\gamma_1\gamma_2\gamma_3}
\;  G^{a\gamma_1, b\gamma_2, c\gamma_3}(x_1 x_2 x_3)\cr
&-\quar\bar\sigma^{\mu\dot\alpha\alpha}
\langle 0| T Q^A_{\alpha 1}  A^{a\alpha_1}(x_1)
\tilde Q_{A\dot\alpha 2}  A^{b\alpha_2}(x_2) A^{c\alpha_3}(x_3) |0\rangle\cr
&-\quar\bar\sigma^{\mu\dot\alpha\alpha}
\langle 0| T \tilde Q_{A\dot\alpha 1}  A^{a\alpha_1}(x_1)
Q^A_{\alpha 2}  A^{b\alpha_2}(x_2) A^{c\alpha_3}(x_3) |0\rangle\cr
&-\quar\bar\sigma^{\mu\dot\alpha\alpha}
\langle 0| T Q^A_{\alpha 1}  A^{a\alpha_1}(x_1)
A^{b\alpha_2}(x_2) \tilde Q_{A\dot\alpha 3} A^{c\alpha_3}(x_3) |0\rangle\cr
&-\quar\bar\sigma^{\mu\dot\alpha\alpha}
\langle 0| T \tilde Q_{A\dot\alpha 1}  A^{a\alpha_1}(x_1)
A^{b\alpha_2}(x_2) Q^A_{\alpha 3}  A^{c\alpha_3}(x_3) |0\rangle\cr
&-\quar\bar\sigma^{\mu\dot\alpha\alpha}
\langle 0| T A^{a\alpha_1}(x_1)
Q^A_{\alpha 2} A^{b\alpha_2}(x_2)
\tilde Q_{A\dot\alpha 3} A^{c\alpha_3}(x_3) |0\rangle\cr
&-\quar  \bar\sigma^{\mu\dot\alpha\alpha}
\langle 0| T A^{a\alpha_1}(x_1)
\tilde Q_{A\dot\alpha 2} A^{b\alpha_2}(x_2) Q^A_{\alpha 3}
A^{c\alpha_3}(x_3) |0\rangle
\label{TQQAAA}\end{align}
$Q^A_\alpha$, $\tilde Q_{A\dot\alpha}$ are the conformal supercharges
appearing in the superconformal group $PSU(2,2|4)$, where
$1\le A\le 4$. In this section
we distinguish spinor indices $1\le \alpha,\dot\alpha\le 2$
from the Lorentz indices $0\le \mu, \alpha_i,\gamma_i\le 3$, and $i$ denotes the
site. The color indices $a,b,c$ run over the dimension of the gauge group.
The notation follows \cite{Haag} where $\sigma^i$ are the Pauli matrices,
\begin{align}
\sigma^\mu_{\alpha\dot\beta} &= (1, \sigma^i),\quad
\bar\sigma^{\mu\;\dot\alpha \beta}= (1,-\sigma^i),
\cr \epsilon_{\alpha\beta} &= - \epsilon_{\beta\alpha},\quad
\epsilon^{\alpha \beta} = - \epsilon^{\beta\alpha},\quad
\epsilon^{12} = \epsilon_{21} = 1\quad
\hbox{same for dotted indices}
\end{align}
The conformal supercharges rotate a gluon into a fermion
in the adjoint representation,
\begin{align}
Q^A_\alpha A^{a\mu} \sim \sigma^\mu_{\alpha\dot\beta} \epsilon^{\dot\beta
\dot\gamma} \bar\psi^{Aa}_{\dot\gamma},\quad
\widetilde Q^A_{\dot\beta}A^{a\mu}
\sim \epsilon_{\dot\beta\dot\alpha}
\bar\sigma^{\mu\;\dot\alpha\gamma} \psi^{Aa}_{\gamma}
\nonumber\end{align}
We start by computing the first $Q\tilde Q$ term in (\ref{TQQAAA}), using
the Lagrangian coupling fermions and gluons
\begin{align}
{\cal L}&= -\quar F^{\mu\nu a} F_{\mu\nu a} - i g\bar\psi^A \bar\sigma^\mu
D_\mu \psi_A
= {\cal L}_0 + {\cal L}_I\cr
{\cal L}_I
&= - g f_{abc} A^b_\mu A^c_\nu \partial^\mu A^{\nu a}
-{g^2\over 4} f_{abc} f_{ade} A^b_\mu A^c_\nu A^{d\mu} A^{e\nu}
-i g f_{def} \bar\psi^{A d}_{\dot\alpha}
\bar\sigma^{\mu\dot\alpha\alpha}
A^e_\mu\psi^f_{A\alpha}
\label{Linter}\end{align}
Then moving from the Heisenberg picture to the interaction picture,
and working to first order in the coupling $g$, and suppressing $g$,
\begin{align}
&-\quar  \bar\sigma^{\mu\dot\alpha\alpha}
\langle 0| T Q^A_{\alpha 1}  A^{a\alpha_1}(x_1)
\tilde Q_{A\dot\alpha 2}  A^{b\alpha_2}(x_2)
A^{c\alpha_3}(x_3) |0\rangle
\cr&= -\quar\bar\sigma^{\mu\dot\alpha\alpha}
\langle 0| T \sigma^{\alpha_1}_{\alpha\dot\beta} \epsilon^{\dot\beta\dot\gamma}
\bar\psi^{Aa}_{\dot\gamma}(x_1) \epsilon_{\dot\alpha\dot\kappa}\bar
\sigma^{\alpha_2\dot\kappa\gamma}\psi^b_{A\gamma}(x_2)
A^{c\alpha_3}(x_3) |0\rangle
\cr &= -\quar\bar\sigma^{\mu\dot\alpha\alpha}\;
\sigma^{\alpha_1}_{\alpha\dot\beta} \epsilon^{\dot\beta\dot\gamma}
\epsilon_{\dot\alpha\dot\kappa}\bar
\sigma^{\alpha_2\dot\kappa\gamma}
\langle 0| T \bar\psi^{Aa}_{\dot\gamma}(x_1)
\psi^{b}_{A\gamma}(x_2)A^{c\alpha_3}(x_3)
e^{i\int d^4 z {\cal L}_I}|0\rangle\cr
&=  -\quar\bar\sigma^{\mu\dot\alpha\alpha}
\sigma^{\alpha_1}_{\alpha\dot\beta} \epsilon^{\dot\beta\dot\gamma}
\epsilon_{\dot\alpha\dot\kappa}\bar
\sigma^{\alpha_2\dot\kappa\gamma} \hskip-5pt\int \hskip-3pt d^4 z \langle 0 |T
\bar\psi^{Aa}_{\dot\gamma}(x_1)
\psi^b_{A\gamma}(x_2)A^{c\alpha_3}(x_3)
\bar\psi^{Bd}_{\dot\delta}(z)
\bar\sigma^{\nu\dot\delta\delta}A^e_\nu (z)
\psi^f_{B\delta}(z)|0\rangle f_{def}\cr
&= \quar \bar\sigma^{\mu\dot\alpha\alpha}\;
\sigma^{\alpha_1}_{\alpha\dot\beta} \epsilon^{\dot\beta\dot\gamma}
\epsilon_{\dot\alpha\dot\kappa}\bar
\sigma^{\alpha_2\dot\kappa\gamma} \bar\sigma^{\nu\dot\delta\delta}
\hskip-3pt\int d^4 z D^{\alpha_3}_\nu(x_3-z) S^F_{\delta\dot\gamma}(z-x_1)
\delta^A_B S^F_{\gamma\dot\delta}(x_2-z) \delta_A^B\;
\delta^{ce}\delta^{af}\delta^{bd} f_{def}
\cr\label{Firstterm}\end{align}
where the gauge and fermion propagators are found from
\begin{align}
&\delta^{ce} D^{\alpha_3}_\nu(x_3-z)=
\langle 0|T A^{c\alpha_3}(x_3) A^e_\nu (z) |0\rangle =
\delta^{ce}\hskip-3pt\int d^4 p_3 e^{-i p_3\cdot (x_3-z)}
\tilde D^{\alpha_3}_\nu (p_3),\;\;
\tilde D^{\alpha_3}_\nu (p) =-{i\over p^2}\cr
& \delta_A^B \delta^{be}
S^F_{\gamma\dot\delta}(x_2-z) = \langle 0|T \psi^b_{A\gamma}(x_2)
\bar\psi^{Be}_{\dot\delta}(z)|0\rangle =\delta_A^B\delta^{be}
\hskip-5pt
\int d^4 p e^{-i p\cdot (x_2-z)} \tilde S^F_{\gamma\dot\delta}(p),
\;\tilde S^F_{\gamma\dot\delta}(p) = {i\sigma^\omega_{\gamma\dot\delta}
p_\omega\over p^2}
\end{align}
Then the Fourier transform of (\ref{Firstterm}) is
\begin{align}
& \bar\sigma^{\mu\dot\alpha\alpha}\;
\sigma^{\alpha_1}_{\alpha\dot\beta} \epsilon^{\dot\beta\dot\gamma}
\epsilon_{\dot\alpha\dot\kappa}\bar
\sigma^{\alpha_2\dot\kappa\gamma} \bar\sigma^{\nu\dot\delta\delta}
\int d^4x_1 d^4x_2 d^4x_3 e^{ik_1\cdot x_1}
e^{ik_2\cdot x_2} e^{ik_3\cdot x_3} \cr
&\hskip40pt \cdot\int d^4 z
D^{\alpha_3}_\nu(x_3-z)  S^F_{\delta\dot\gamma}(z-x_1)
S^F_{\gamma\dot\delta}(x_2-z)f_{abc}\cr
&=\bar\sigma^{\mu\dot\alpha\alpha}\;
\sigma^{\alpha_1}_{\alpha\dot\beta} \epsilon^{\dot\beta\dot\gamma}
\epsilon_{\dot\alpha\dot\kappa}\bar
\sigma^{\alpha_2\dot\kappa\gamma} \bar\sigma^{\nu\dot\delta\delta}
\;(2\pi)^4\delta^4(k_1+k_2+k_3) \tilde S^F_{\delta\dot\gamma}
(-k_1)\tilde S^F_{\gamma\dot\delta}(k_2) \tilde D^{\alpha_3}_\nu (k_3)
\;f_{abc}\cr
&= \;(2\pi)^4\delta^4(k_1+k_2+k_3)\;
\bar\sigma^{\mu\dot\alpha\alpha}\;
\sigma^{\alpha_1}_{\alpha\dot\beta} \epsilon^{\dot\beta\dot\gamma}
\epsilon_{\dot\alpha\dot\kappa}\bar
\sigma^{\alpha_2\dot\kappa\gamma} \bar\sigma^{\nu\dot\delta\delta}\;
{(-i\sigma^\omega_{\delta\dot\gamma} k_{1\omega})\over k_1^2}
{(i\sigma^\zeta_{\gamma\dot\delta} k_{2\zeta})\over k_2^2}
\tilde D^{\alpha_3}_\nu (k_3)\; f_{abc}\cr
\label{qbarqft}\end{align}
Since the three-gluon amplitude in (\ref{threeg}) has had the
three external propagators truncated, we now truncate
(\ref{qbarqft}) by multiplying it by the inverse propagators,
$\tilde D^{-1\;\tilde\alpha_1\alpha_1}(k_1)
\tilde  D^{-1\;\tilde\alpha_2\alpha_2}(k_2)
\tilde D^{-1\;\tilde\alpha_3\alpha_3}(k_3)
= -\eta^{\tilde\alpha_1\alpha_1} k_1^2\eta^{\tilde\alpha_2\alpha_2}k_2^2
\tilde D^{-1\;\tilde\alpha_3\alpha_3}(k_3)$,
where $\tilde D^{-1\;\tilde\alpha\alpha}(k) \tilde D_{\alpha\nu}(k)
= \delta^{\tilde\alpha}_\nu$. Then  (\ref{qbarqft}) truncated becomes
\begin{align}
&- \;(2\pi)^4\delta^4(k_1+k_2+k_3)\;
\bar\sigma^{\mu\dot\alpha\alpha}\;
\sigma^{\tilde\alpha_1}_{\alpha\dot\beta} \epsilon^{\dot\beta\dot\gamma}
\epsilon_{\dot\alpha\dot\kappa}\bar
\sigma^{\tilde\alpha_2\dot\kappa\gamma} \bar\sigma^{\tilde\alpha_3
\dot\delta\delta}\; \sigma^\omega_{\delta\dot\gamma}
\sigma^\zeta_{\gamma\dot\delta} k_{1\omega} k_{2\zeta} \;f_{abc}\cr
&= - \;(2\pi)^4\delta^4(k_1+k_2+k_3)\; (-(\bar\sigma^{\tilde\alpha_2}
\sigma^\zeta\bar\sigma^{\tilde\alpha_3}\sigma^\omega\bar\sigma^{\tilde\alpha_1}
\sigma^\mu)^{\dot\kappa}_{\hskip5pt \dot\kappa})\;
k_{1\omega}k_{2\zeta} \;f_{abc}
\label{TQQA1}\end{align}
Here we have used properties of the $\sigma, \bar\sigma$ matrices
\begin{align}
\epsilon^{\tau\kappa} \sigma^\mu_{\kappa\dot\beta}
\epsilon^{\dot\beta\dot\gamma}
= - (\sigma^2\sigma^\mu\sigma^2)^{\tau\dot\gamma}
= - \bar\sigma^{\mu\dot\gamma\tau}
\end{align}
since $\sigma^2\sigma^\mu\sigma^2 = \bar\sigma^{\mu T}$.
And similarly $\epsilon_{\dot\kappa\dot\alpha} \bar\sigma^{\mu\dot\alpha\alpha}
\epsilon_{\alpha\tau} = -(\sigma^2\bar\sigma\sigma^2)_{\dot\kappa\tau}
= - \sigma^\mu_{\tau\dot\kappa}$, since $\sigma^2\bar\sigma^\mu\sigma^2
= \sigma^{\mu T}.$
The second $Q\tilde Q$ term in (\ref{TQQAAA}) leads to
\begin{align}
&-\quar\bar\sigma^{\mu\dot\alpha\alpha}
\langle 0| T \tilde Q_{A\dot\alpha 1}  A^{a\alpha_1}(x_1)
Q^A_{\alpha 2}  A^{b\alpha_2}(x_2)
A^{c\alpha_3}(x_3) |0\rangle
\cr &\rightarrow
- \;(2\pi)^4\delta^4(k_1+k_2+k_3)\; (\sigma^\mu\bar\sigma^{\tilde\alpha_1}
\sigma^\omega\bar\sigma^{\tilde\alpha_3}\sigma^\zeta\bar\sigma^{\tilde\alpha_2}
)_\kappa^{\hskip5pt\kappa}\;k_{1\omega}k_{2\zeta}\;
f_{acb}
\label{TQQA2}\end{align}
On combining (\ref{TQQA1}) and (\ref{TQQA2}),
the truncated Fourier transform sums to
\begin{align}
&-\quar\bar\sigma^{\mu\dot\alpha\alpha}
\langle 0| T Q^A_{\alpha 1}  A^{a\alpha_1}(x_1)
\tilde Q_{A\dot\alpha 2}  A^{b\alpha_2}(x_2)
A^{c\alpha_3}(x_3) |0\rangle
\cr &-\quar  \bar\sigma^{\mu\dot\alpha\alpha}
\langle 0| T \tilde Q_{A\dot\alpha 1}  A^{a\alpha_1}(x_1)
Q^A_{\alpha 2}  A^{b\alpha_2}(x_2)
A^{c\alpha_3}(x_3) |0\rangle\cr
&\rightarrow
 \;(2\pi)^4\delta^4(k_1+k_2+k_3)\; \Big((\bar\sigma^{\tilde\alpha_2}
\sigma^\zeta\bar\sigma^{\tilde\alpha_3}\sigma^\omega\bar\sigma^{\tilde\alpha_1}
\sigma^\mu)^{\dot\kappa}_{\hskip5pt \dot\kappa}
+  (\sigma^\mu\bar\sigma^{\tilde\alpha_1}
\sigma^\omega\bar\sigma^{\tilde\alpha_3}
\sigma^\zeta\bar\sigma^{\tilde\alpha_2})_\kappa^{\hskip5pt\kappa}\Big)\;
k_{1\omega}k_{2\zeta}\; f_{abc}\cr
&=  \;(2\pi)^4\delta^4(k_1+k_2+k_3)\;
tr (\gamma^{\tilde\alpha_2}\gamma^\zeta\gamma^{\tilde\alpha_3}
\gamma^\omega\gamma^{\tilde\alpha_1}\gamma^\mu)
\; k_{1\omega}k_{2\zeta}\; f_{abc}
\label{comb12}\end{align}
Here we use further properties of the $\gamma$ matrices in the Weyl
representation (\ref{gm1})
\begin{align}
&\gamma^\mu\gamma^\nu\gamma^\rho\gamma^\lambda\gamma^\omega\gamma^\zeta
=\left(\begin{matrix} (\sigma^\mu\bar\sigma^\nu\sigma^\rho\bar\sigma^\lambda
\sigma^\omega\bar\sigma^\zeta)_\kappa^{\hskip3pt\tau}&0\cr
0& (\bar \sigma^\mu\sigma^\nu\bar\sigma^\rho\sigma^\lambda
\bar\sigma^\omega\sigma^\zeta)^{\dot\kappa}_{\hskip3pt\dot\tau}
\end{matrix}\right),
\cr&tr (\gamma^\mu\gamma^\nu\gamma^\rho\gamma^\lambda\gamma^\omega\gamma^\zeta)
\cr&= (\sigma^\mu\bar\sigma^\nu\sigma^\rho\bar\sigma^\lambda
\sigma^\omega\bar\sigma^\zeta)_\kappa^{\hskip3pt\kappa}
+ (\bar \sigma^\mu\sigma^\nu\bar\sigma^\rho\sigma^\lambda
\bar\sigma^\omega\sigma^\zeta)^{\dot\kappa}_{\hskip3pt\dot\kappa}
= (\sigma^\zeta\bar\sigma^\omega\sigma^\lambda\bar\sigma^\rho
\sigma^\nu\bar\sigma^\mu)_\kappa^{\hskip3pt\kappa}
+ (\bar \sigma^\mu\sigma^\nu\bar\sigma^\rho\sigma^\lambda
\bar\sigma^\omega\sigma^\zeta)^{\dot\kappa}_{\hskip3pt\dot\kappa}
\label{gammatr2}\cr\end{align}
The contribution of the last four terms of (\ref{TQQAAA}) can be
found by exchanging $b,\alpha_2,k_2\rightarrow c,\alpha_3, k_3$
and from there $a,\alpha_1,k_1\rightarrow b,\alpha_2, k_2$
by inspection, to yield that
the truncated Fourier transform
of the last six terms of (\ref{TQQAAA}) is
\begin{align}
&-\quar\bar\sigma^{\mu\dot\alpha\alpha}
\langle 0| T Q^A_{\alpha 1}  A^{a\alpha_1}(x_1)
\tilde Q_{A\dot\alpha 2}  A^{b\alpha_2}(x_2)
A^{c\alpha_3}(x_3) |0\rangle
\cr&-\quar\bar\sigma^{\mu\dot\alpha\alpha}
\langle 0| T \tilde Q_{A\dot\alpha 1}  A^{a\alpha_1}(x_1)
Q^A_{\alpha 2}  A^{b\alpha_2}(x_2)
A^{c\alpha_3}(x_3) |0\rangle\cr
&-\quar\bar\sigma^{\mu\dot\alpha\alpha}
\langle 0| T Q^A_{\alpha 1}  A^{a\alpha_1}(x_1)
A^{b\alpha_2}(x_2)
\tilde Q_{A\dot\alpha 3} A^{c\alpha_3}(x_3) |0\rangle
\cr&-\quar\bar\sigma^{\mu\dot\alpha\alpha}
\langle 0| T \tilde Q_{A\dot\alpha 1}  A^{a\alpha_1}(x_1)
A^{b\alpha_2}(x_2)
Q^A_{\alpha 3}  A^{c\alpha_3}(x_3) |0\rangle\cr
&-\quar\bar\sigma^{\mu\dot\alpha\alpha}
\langle 0| T A^{a\alpha_1}(x_1)
Q^A_{\alpha 2} A^{b\alpha_2}(x_2)
\tilde Q_{A\dot\alpha 3} A^{c\alpha_3}(x_3) |0\rangle
\cr&-\quar\bar\sigma^{\mu\dot\alpha\alpha}
\langle 0| T A^{a\alpha_1}(x_1)
\tilde Q_{A\dot\alpha 2} A^{b\alpha_2}(x_2) Q^A_{\alpha 3}
A^{c\alpha_3}(x_3) |0\rangle\label{QAQAA}\\
&\rightarrow
\;(2\pi)^4\delta^4(k_1+k_2+k_3)
\Big(
tr (\gamma^{\tilde\alpha_2}\gamma^\zeta\gamma^{\tilde\alpha_3}
\gamma^\omega\gamma^{\tilde\alpha_1}\gamma^\mu)\; k_{1\omega}k_{2\zeta}\;
\cr&\hskip90pt
- tr(\gamma^{\tilde\alpha_3}\gamma^\zeta\gamma^{\tilde\alpha_2}
\gamma^\omega\gamma^{\tilde\alpha_1}\gamma^\mu)\;
k_{1\omega} k_{3\zeta}\;
+tr(\gamma^{\tilde\alpha_3}
\gamma^\zeta\gamma^{\tilde\alpha_1}\gamma^\omega\gamma^{\tilde\alpha_2}
\gamma^\mu)\;  k_{2\omega}k_{3\zeta}
\Big) {f_{abc}}\label{trtr}
\cr\end{align}
which with the coupling $g$ reinserted, cancels (\ref{PA3f}).
This motivates the identity (\ref{pmub}).

In summary, we have observed that the action of the $SO(2,4)$ Yangian
on a pure gluon amplitude is equivalent to an appropriately truncated
amplitude with
2 fermions and two less gluons, as expected
from the $PSU(2,2|4)$ Yangian symmetry of the pure gluon amplitude.
If we could interpret the fermion amplitude as the effect of
some process in pure Yang-Mills theory, we might
realize a role for the level one generators of  the
$SO(2,4)$ Yangian in pure gluon theory. Such an analysis could
lead to a deeper understanding of the dynamics of non-supersymmetric
non-abelian gauge theory, and help to interpret a non-supersymmetric
extension of \cite{AH, Beisert}. 

We give an analogous identity for the four-point partial gluon amplitude,
whose derivation illuminates further how to extend these identities for all $N$. 
\vskip20pt

The on-shell four-gluon tree amplitude \cite{BCJ} is
\begin{align}
&A^{a\alpha_1,\; b\alpha_2,\; c\alpha_3,\;d\alpha_4}
=  g^2 \,(2\pi)^4\delta^4(k_1+k_2+k_3+k_4)\;
\cr &\hskip 90pt\cdot
\big( \;{c_s n^{\alpha_1\alpha_2\alpha_3\alpha_4}_s\over s}
+ {c_t n_t^{\alpha_1\alpha_2\alpha_3\alpha_4}\over t}
+ {c_u n^{\alpha_1\alpha_2\alpha_3\alpha_4}_u\over u}\;\big)
\nonumber\end{align}
\begin{align} c_s \equiv &f_{abe} f_{ecd}, \quad
c_t \equiv f_{bce} f_{ead}, \quad
c_u \equiv f_{cae} f_{ebd}, \quad c_s + c_u + c_t = 0,\quad
n_s + n_u + n_t = 0
\end{align}
where
$s=s_{12}, t=s_{23}, u= (k_1+k_3)^2$.
We focus on the gauge invariant partial amplitude $A(1234)$,
where the polarization vectors have been removed,
\begin{align}
&A(1234)= g^2  \,(2\pi)^4\delta^4(\sum k_i)\;
\big( \;{in_s\over s}- {in_t\over t} \;\big),\cr\cr
&i n_s^{\alpha_1\alpha_2\alpha_3\alpha_4}\cr &=
\big(\eta^{\alpha_1\alpha_2}(k_1-k_2)_\sigma
+ 2 k_2^{\alpha_1}\delta^{\alpha_2}_\sigma -2k_1^{\alpha_2}
\delta^{\alpha_1}_{\sigma}\big)
\big(\eta^{\alpha_3\alpha_4}(k_3-k_4)^\sigma
+ 2k_4^{\alpha_3}\eta^{\sigma\alpha_4}
-2 k_3^{\alpha_4}\eta^{\sigma\alpha_3}\big)\cr
&\hskip7pt + \big(\eta^{\alpha_1\alpha_3}\eta^{\alpha_2\alpha_4}
- (\eta^{\alpha_1\alpha_4}\eta^{\alpha_2\alpha_3}\big)\; s\cr
\cr &i n_t^{\alpha_1\alpha_2\alpha_3\alpha_4} \cr&=
\big(\eta^{\alpha_2\alpha_3}(k_2-k_3)_\sigma
+ 2 k_2^{\alpha_3}\delta^{\alpha_3}_\sigma -2k_2^{\alpha_3}
\delta^{\alpha_2}_{\sigma}\big)
\big(\eta^{\alpha_4\alpha_1}(k_1-k_4)^\sigma
- 2k_1^{\alpha_4}\eta^{\sigma\alpha_1}
+2 k_4^{\alpha_1}\eta^{\sigma\alpha_4}\big)\cr
&\hskip7pt + \big(-\eta^{\alpha_2\alpha_4}\eta^{\alpha_3\alpha_1}
+ (\eta^{\alpha_2\alpha_1}\eta^{\alpha_3\alpha_4}\big)\; t\cr
\label{nsnt}\end{align}
Since $\hat P^\mu$ commutes with $\delta^4(\sum k_i)$,
in Appendix B we derive
\begin{align}
&-\hat P^\mu\;  \Big( {in_s\over s} - {in_t\over t}\Big)\cr &=
\Bigg[
\eta^{\mu\alpha_1}\Big(\eta^{\alpha_2\alpha_3} \big(
4k_3^{\alpha_4} - {8 k_1^{\alpha_4} k_1\cdot k_2\over t}\big)
-\eta^{\alpha_3\alpha_4} (4k_3^{\alpha_2}
- {8 k_1^{\alpha_2} k_1\cdot k_4\over s}\big)
\cr&\hskip48pt
-{8 k_1^{\alpha_2} \; (k_1^{\alpha_4} k_4^{\alpha_3}
+ k_3^{\alpha_4} k_2^{\alpha_3})\over s}
+{8 k_1^{\alpha_4} \; (k_1^{\alpha_2} k_2^{\alpha_3}
+ k_3^{\alpha_2} k_4^{\alpha_3})\over t}\; \Big)\cr
&\hskip15pt
+ \eta^{\mu\alpha_2}
\Big(\eta^{\alpha_1\alpha_4} \big(
-4k_4^{\alpha_3} + {8 k_2^{\alpha_3} k_1\cdot k_2\over t}\big)
+\eta^{\alpha_3\alpha_4} (4k_4^{\alpha_1}
- {8 k_2^{\alpha_1} k_1\cdot k_4\over s}\big)
\cr&\hskip48pt
+{8 k_2^{\alpha_1} \; (k_2^{\alpha_3} k_3^{\alpha_4}
+ k_4^{\alpha_3} k_1^{\alpha_4})\over s}
-{8 k_2^{\alpha_3} \; (k_2^{\alpha_1} k_1^{\alpha_4}
+ k_4^{\alpha_1} k_3^{\alpha_4})\over t}\; \Big)\cr
&\hskip15pt + \;k_1^\mu\;\Big(
6 \;\eta^{\alpha_1\alpha_3}\eta^{\alpha_2\alpha_4}
-\eta^{\alpha_1\alpha_4}\eta^{\alpha_2\alpha_3}\;
\big(6+{8k_1\cdot k_2\over t})
-\eta^{\alpha_1\alpha_2}\eta^{\alpha_3\alpha_4}\;
\big(6+{16 k_1\cdot k_4\over s})\cr
&\hskip48pt + \eta^{\alpha_1\alpha_2}\;
\;\big( {16 k_4^{\alpha_3} k_1^{\alpha_4} -
16 k_3^{\alpha_4} k_1^{\alpha_3} - 4 k_4^{\alpha_3} k_3^{\alpha_4}\over s}
- {12 k_1^{\alpha_4} k_2^{\alpha_3}\over t}\big)\cr
&\hskip48pt + \eta^{\alpha_3\alpha_4}\;
\;\big( {8 k_1^{\alpha_2} k_4^{\alpha_1} -
16 k_2^{\alpha_1} k_4^{\alpha_2} - 4 k_1^{\alpha_2} k_2^{\alpha_1}\over s}
- {12 k_3^{\alpha_2} k_4^{\alpha_1}\over t}\big)\cr
&\hskip48pt + \eta^{\alpha_1\alpha_3}\;
\;\big( {8 k_1^{\alpha_2} k_3^{\alpha_4} \over s}
+ {12 k_1^{\alpha_4} k_3^{\alpha_2}\over t}\big)
+ \eta^{\alpha_2\alpha_4}\;
\;\big( {16 k_2^{\alpha_1} k_4^{\alpha_3} \over s}
+ {12 k_2^{\alpha_3} k_4^{\alpha_1}\over t}\big)
\cr&\hskip48pt + \eta^{\alpha_1\alpha_4}\;
\;\big( -{8 k_1^{\alpha_2} k_4^{\alpha_3}\over s}
+ {-8 k_2^{\alpha_3} k_4^{\alpha_2}
+8 k_3^{\alpha_2} k_4^{\alpha_3} + 4k_2^{\alpha_3} k_3^{\alpha_2}\over t}
\big)\cr
&\hskip48pt + \eta^{\alpha_2\alpha_3}\;
\;\big( -{16 k_2^{\alpha_1} k_3^{\alpha_4}\over s}
+ {-12 k_1^{\alpha_4} k_3^{\alpha_1}
+ 12 k_4^{\alpha_1} k_3^{\alpha_4} \over t}\big)\;\Big)
\cr
&\hskip15pt
+ k_2^\mu\;\Big(
2 \;\eta^{\alpha_1\alpha_3}\eta^{\alpha_2\alpha_4}
-\eta^{\alpha_1\alpha_4}\eta^{\alpha_2\alpha_3}\;
\big(2+{8k_1\cdot k_2\over t})
- 2\eta^{\alpha_1\alpha_2}\eta^{\alpha_3\alpha_4}\cr
&\hskip48pt + \eta^{\alpha_1\alpha_2}\;
\;\big( {4 k_4^{\alpha_3} k_3^{\alpha_4} \over s}
+ {4 k_1^{\alpha_4} k_2^{\alpha_3}\over t}\big)
+ \eta^{\alpha_3\alpha_4}\;
\;\big( {4 k_2^{\alpha_1} (k_3-k_4)^{\alpha_2}\over s}
- {12 k_3^{\alpha_2} k_4^{\alpha_1}\over t}\big)\cr
&\hskip48pt + \eta^{\alpha_1\alpha_3}\;
\;\big( {4 k_1^{\alpha_4} k_3^{\alpha_2} \over t}\big)
+ \eta^{\alpha_2\alpha_4}\;
\;\big( {8 k_2^{\alpha_1} k_4^{\alpha_3} \over s}
+ {4 k_2^{\alpha_3} k_4^{\alpha_1}\over t}\big)
+ \eta^{\alpha_1\alpha_4}\;
\;\big( { 4 k_3^{\alpha_2} (k_4 - k_1)^{\alpha_3}\over t}\big)\cr
&\hskip48pt + \eta^{\alpha_2\alpha_3}\;
\;\big(
{-8 k_2^{\alpha_1} k_3^{\alpha_4}\over s} +
{4 k_1^{\alpha_4} k_2^{\alpha_1} - 4 k_4^{\alpha_1} k_2^{\alpha_4}
+ 8 k_3^{\alpha_4} k_4^{\alpha_1}\over t}
\big)\;\Big)\cr
&\hskip20pt - (1\leftrightarrow 4), (2\leftrightarrow 3)
\;\hbox{simultaneous exchange}\;\Bigg]
\label{summary1}\end{align}
which can also be expressed in terms of Dirac matrices
\begin{align}
&={(k_1+k_2)_\tau\over (k_1+k_2)^2}\;
\Big[ -tr (\gamma^{\tilde\alpha_1}\gamma^\mu\gamma^{\tilde\alpha_3}
\gamma^\zeta\gamma^{\tilde\alpha_4}\gamma^\tau\gamma^{\tilde
\alpha_2}\gamma^\omega) k_{1\omega}k_{3\zeta}
+ tr (\gamma^{\tilde\alpha_2}\gamma^\mu\gamma^{\tilde\alpha_3}
\gamma^\zeta\gamma^{\tilde\alpha_4}\gamma^\tau\gamma^{\tilde
\alpha_1}\gamma^\omega) k_{2\omega}k_{3\zeta}\cr
&\hskip 60pt + tr (\gamma^{\tilde\alpha_1}\gamma^\mu\gamma^{\tilde\alpha_4}
\gamma^\zeta\gamma^{\tilde\alpha_3}\gamma^\tau\gamma^{\tilde
\alpha_2}\gamma^\omega) k_{1\omega}k_{4\zeta}
- tr (\gamma^{\tilde\alpha_2}\gamma^\mu\gamma^{\tilde\alpha_4}
\gamma^\zeta\gamma^{\tilde\alpha_3}\gamma^\tau\gamma^{\tilde
\alpha_1}\gamma^\omega) k_{2\omega}k_{4\zeta}\Big]\cr
&-{(k_2+k_3)_\tau\over (k_2+k_3)^2}\;
\Big[ tr (\gamma^{\tilde\alpha_1}\gamma^\mu\gamma^{\tilde\alpha_2}
\gamma^\zeta\gamma^{\tilde\alpha_3}\gamma^\tau\gamma^{\tilde
\alpha_4}\gamma^\omega) k_{1\omega}k_{2\zeta}
- tr (\gamma^{\tilde\alpha_1}\gamma^\mu\gamma^{\tilde\alpha_3}
\gamma^\zeta\gamma^{\tilde\alpha_2}\gamma^\tau\gamma^{\tilde
\alpha_4}\gamma^\omega) k_{1\omega}k_{3\zeta} \cr
&\hskip 60pt + tr (\gamma^{\tilde\alpha_2}\gamma^\mu\gamma^{\tilde\alpha_4}
\gamma^\zeta\gamma^{\tilde\alpha_1}\gamma^\tau\gamma^{\tilde
\alpha_3}\gamma^\omega) k_{2\omega}k_{4\zeta}
- tr (\gamma^{\tilde\alpha_3}\gamma^\mu\gamma^{\tilde\alpha_4}
\gamma^\zeta\gamma^{\tilde\alpha_1}\gamma^\tau\gamma^{\tilde
\alpha_2}\gamma^\omega) k_{3\omega}k_{4\zeta}\Big] \cr
&-\Bigg[{k_{1\omega} k_{2\tau}\over (k_3+k_4)^2}
\; \Big[ -2 k_4^{\tilde\alpha_3}\;
tr (\gamma^{\tilde\alpha_1}\gamma^\mu\gamma^{\tilde\alpha_2}
\gamma^\tau\gamma^{\tilde\alpha_4}\gamma^\omega) + 2k_3^{\tilde\alpha_4}
tr (\gamma^{\tilde\alpha_1}\gamma^\mu\gamma^{\tilde\alpha_2}
\gamma^\tau\gamma^{\tilde\alpha_3}\gamma^\omega)
\cr&\hskip200pt +  \eta^{\tilde\alpha_3\tilde\alpha_4}
\Big(-k_3+k_4 \Big)_\nu\;
tr (\gamma^{\tilde\alpha_1}\gamma^\mu\gamma^{\tilde\alpha_2}
\gamma^\tau\gamma^\nu\gamma^\omega)\Big]\cr
&\hskip30pt  + {k_{3\omega} k_{4\tau}\over (k_3+k_4)^2}
\Big[-2 k_2^{\tilde\alpha_1}\;
tr (\gamma^{\tilde\alpha_3}\gamma^\mu\gamma^{\tilde\alpha_4}
\gamma^\tau\gamma^{\tilde\alpha_2}\gamma^\omega)
+ 2k_1^{\tilde\alpha_2}
tr (\gamma^{\tilde\alpha_3}\gamma^\mu\gamma^{\tilde\alpha_4}
\gamma^\tau\gamma^{\tilde\alpha_1}\gamma^\omega)
\cr&\hskip 200pt +  \eta^{\tilde\alpha_1\tilde\alpha_2}
\Big(-k_1+k_2 \Big)_\nu\;
tr (\gamma^{\tilde\alpha_3}\gamma^\mu\gamma^{\tilde\alpha_4}
\gamma^\tau\gamma^\nu\gamma^\omega)
\Big]\Bigg]\cr
&+\;\Bigg[{k_{2\omega} k_{3\tau}\over (k_1+k_4)^2}
\; \Big[ -2 k_4^{\tilde\alpha_1}\;
tr (\gamma^{\tilde\alpha_2}\gamma^\mu\gamma^{\tilde\alpha_3}
\gamma^\tau\gamma^{\tilde\alpha_4}\gamma^\omega)
+ 2k_1^{\tilde\alpha_4}
\;tr (\gamma^{\tilde\alpha_2}\gamma^\mu\gamma^{\tilde\alpha_3}
\gamma^\tau\gamma^{\tilde\alpha_1}\gamma^\omega)
\cr&\hskip200pt +  \eta^{\tilde\alpha_1\tilde\alpha_4}
\Big(-k_1+k_4 \Big)_\nu\;
tr (\gamma^{\tilde\alpha_2}\gamma^\mu\gamma^{\tilde\alpha_3}
\gamma^\tau\gamma^\nu\gamma^\omega)\Big]\cr
&\hskip30pt + {k_{1\omega} k_{4\tau}\over (k_2+k_3)^2}
\Big[-2 k_3^{\tilde\alpha_2}\;
tr (\gamma^{\tilde\alpha_1}\gamma^\mu\gamma^{\tilde\alpha_4}
\gamma^\tau\gamma^{\tilde\alpha_3}\gamma^\omega)
+ 2k_2^{\tilde\alpha_3}
tr (\gamma^{\tilde\alpha_1}\gamma^\mu\gamma^{\tilde\alpha_4}
\gamma^\tau\gamma^{\tilde\alpha_2}\gamma^\omega)
\cr&\hskip 200pt
+  \eta^{\tilde\alpha_2\tilde\alpha_3}
\Big(-k_2+k_3 \Big)_\nu\;
tr (\gamma^{\tilde\alpha_1}\gamma^\mu\gamma^{\tilde\alpha_4}
\gamma^\tau\gamma^\nu\gamma^\omega)
\Big]\Bigg]\cr
\label{QQQQ}\end{align}
The equivalence between (\ref{summary1}) and (\ref{QQQQ}) is motivated in
Appendix C, and can be checked manually via the trace relations
(\ref{stantr}),
or by using computer algebra.
Here we have isolated a supercharge contribution (\ref{QQQQ})
for a Yangian level one generator without using the spinor helicity
and superspace formalism.
We display explicity that the $SO(2,4)$ Yangian level one
generator acts on the four-point pure gluon amplitude by transforming it
to another amplitude which could be interpreted to involve fermions,
fields not in the pure gauge theory.
Also the differential operator representation provides more versatility
when comparing how the Yangian acts on scalar and gauge theories.

\section*{Acknowledgements}
\label{Ack}

ND was partially supported by the Hamilton and Silver Awards, provided by 
the Department of Physics and Astronomy of the University of North Carolina 
at Chapel Hill, during the summers of 2021 and 2022. 
LD and ND were partially supported by the UNC Bahnson fund. 

\begin{appendices}
\section{Commutation with the Delta Function}
\label{commutedelta}

We show how $\hat P^\mu$
commutes with the delta function
\begin{align}
\hat P^\mu \; \delta^n (\sum_{i=1}^N k_i)\; M(k)&=
\delta^n (\sum_{i=1}^N k_i) \; \hat P^\mu \; M(k)
\end{align}
for any function of the momenta $M(k)= M(k_1,\ldots, k_N)$ as follows.
\begin{align}
&-\hat P^\mu \; \delta^n (\sum_{i=1}^N k_i)\; M(k)\cr&=
\sum_{1\le i<j\le N} \;\Big[
P^\mu_i D_j +  P_{\rho i} L_j^{\mu\rho} - (i\leftrightarrow j)\Big]
\;   \delta^n (\sum_{i=1}^N k_i)\; M(k)\cr
&= \sum_{1\le i<j\le N} \;\Big[
k^\mu_i (d + k_j\cdot \partial_j)  +  k_{i\rho}
(k_j^\mu\partial_j^\rho - k_j^\rho\partial_j^\mu + \Sigma^{\mu\rho}_j)
- (i\leftrightarrow j)\Big]
\;   \delta^n (\sum_{i=1}^N k_i)\; M(k)\cr
&=\delta^n (\sum_{i=1}^N k_i)\; (-\hat P^\mu)
\;   M(k)\cr
&\hskip7pt + 2
\Bigg( \sum_{1\le i<j\le N} \;\Big[
k^\mu_i k_j\cdot \partial_j  +  k_{i\rho}
(k_j^\mu\partial_j^\rho - k_j^\rho\partial_j^\mu) - (i\leftrightarrow j)\Big]
\;   \delta^n (\sum_{i=1}^N k_i)\;\Bigg)\; M(k)\cr
\end{align}
The last term is zero because the delta function is a function of
the sum of the momentum $\sum_{i=1}^N k_i$, which implies that all the
partial derivatives $\partial_i$ act on it in the same way,
$\partial_i\rightarrow {\partial\over \partial \sum_{\ell=1}^N k_\ell}$,
\begin{align}
& 
\sum_{1\le i<j\le N} \;\Big[
k^\mu_i k_j\cdot \partial_j  +  k_{i\rho}
(k_j^\mu\partial_j^\rho - k_j^\rho\partial_j^\mu) - (i\leftrightarrow j)\Big]
\; \delta^n (\sum_{i=1}^N k_i)\;\cr
&=  \sum_{1\le i<j\le N} \;\Big[ (k_i^\mu k_{j\rho}
+ k_j^\mu k_{i\rho})\; \partial_j^\rho
- k_{i\rho} k_j^\rho \partial_j^\mu - (i\leftrightarrow j)\Big]
\; \delta^n (\sum_{i=1}^N k_i)\cr
&=  \sum_{1\le i<j\le N} \;\Big[ (k_i^\mu k_{j\rho}
+ k_j^\mu k_{i\rho})\; {\partial\over \partial \sum_{\ell=1}^N k_{\ell\rho}}
- k_{i\rho} k_j^\rho
\; {\partial\over \partial \sum_{\ell=1}^N k_{\ell\mu}}
- (i\leftrightarrow j)\Big]
\; \delta^n (\sum_{i=1}^N k_i) =0\cr
\end{align}
since the coefficients of
$\; {\partial\over \partial \sum_{\ell=1}^N k_{\ell\rho}}$ or
$\; {\partial\over \partial \sum_{\ell=1}^N k_{\ell\mu}}$
sum to zero.

\section{\bf A Level One Generator  on the Four-Point Gauge Amplitude }
\label{hatPG4}

We simplify the expression for $\hat P^\mu$ using momentum conservation
and $k_i^2=0$. For $N=4$, $d=1$,
\begin{align}
-\hat P^{\mu\;\alpha_1\alpha_2\alpha_3\alpha_4}_{\gamma_1\gamma_2\gamma_3
\gamma_4} &=\sum_{1\;\le i<j\le 4}\Big( P_i^\mu D_j + P_{i\rho}L_j^{\mu\rho}
- (i\leftrightarrow j)
\Big)^{\alpha_1\alpha_2\alpha_3\alpha_4}_{\;\gamma_1\gamma_2\gamma_3\gamma_4}
\cr&=
\Big(\;(3 k_1 +k_2 -k_3 -3k_4)^\mu \cr
&\hskip15pt  +2 k_1^\mu k_1\cdot\partial_1
+ 2 (k_1+k_2)^\mu k_2\cdot\partial_2
- 2 (k_3+k_4)^\mu k_3\cdot\partial_3
- 2k_4^\mu k_4\cdot\partial_4\cr
&\hskip15pt
+ 2k_2^\mu k_1\cdot \partial_2 - 2k_3^\mu k_4\cdot \partial_3
- 2 k_1\cdot k_2 \partial^\mu_2 + 2 k_1\cdot k_2 \partial^\mu_3\Big)
\; \delta^{\alpha_1}_{\gamma_1} \delta^{\alpha_2}_{\gamma_2}
\delta^{\alpha_3}_{\gamma_3}\delta^{\alpha_4}_{\gamma_4} \cr
&\hskip15pt +\eta^{\mu\alpha_1} k_{1\gamma}  \; \delta^{\alpha_2}_{\gamma_2}
\delta^{\alpha_3}_{\gamma_3}\delta^{\alpha_4}_{\gamma_4}
+ \eta^{\mu\alpha_2} (2k_1 +k_2)_{\gamma_2}  \; \delta^{\alpha_1}_{\gamma_1}
\delta^{\alpha_3}_{\gamma_3}\delta^{\alpha_4}_{\gamma_4}\cr
&\hskip15pt -\eta^{\mu\alpha_3} (k_3+2k_4)_{\gamma_3}
\; \delta^{\alpha_1}_{\gamma_1}
\delta^{\alpha_2}_{\gamma_2}\delta^{\alpha_4}_{\gamma_4}
- \eta^{\mu\alpha_4} k_{4\gamma_4}\; \delta^{\alpha_1}_{\gamma_1}
\delta^{\alpha_2}_{\gamma_2}\delta^{\alpha_3}_{\gamma_3}\cr
&\hskip15pt -\delta^\mu_{\gamma_2} 2 k_1^{\alpha_2}
\delta^{\alpha_1}_{\gamma_1}
\delta^{\alpha_3}_{\gamma_3}\delta^{\alpha_4}_{\gamma_4}
+ \delta^\mu_{\gamma_3} 2 k_4^{\alpha_3}
\delta^{\alpha_1}_{\gamma_1}
\delta^{\alpha_2}_{\gamma_2}\delta^{\alpha_4}_{\gamma_4}
\label{psim}\end{align}
Then from (\ref{nsnt}) and (\ref{prePs}),
\begin{align}
&-\hat P^\mu \; i {n_s\over s} \cr&=
\sum_{1\;\le i<j\le 4}\Big( P_i^\mu D_j + P_{i\rho}L_j^{\mu\rho}
- (i\leftrightarrow j)
\Big)^{\alpha_1\alpha_2\alpha_3\alpha_4}_{\;\gamma_1\gamma_2\gamma_3\gamma_4}
\;\;  {i\over s} \; n_s^{\gamma_1\gamma_2\gamma_3\gamma_4}\cr
&= \sum_{1\;\le i<j\le 4}\Big( P_i^\mu D_j + P_{i\rho}L_j^{\mu\rho}
- (i\leftrightarrow j)
\Big)^{\alpha_1\alpha_2\alpha_3\alpha_4}_{\;\gamma_1\gamma_2\gamma_3\gamma_4}
\;\;(\eta^{\gamma_1\gamma_3}\eta^{\gamma_2\gamma_4}-
\eta^{\gamma_1\gamma_4}\eta^{\gamma_2\gamma_3})\cr
&+ 2(k_3+k_4)^\mu\;\;  {1\over s} \;
\big(\eta^{\alpha_1\alpha_2}(k_1-k_2)_\sigma
+ 2 k_2^{\alpha_1}\delta^{\alpha_2}_\sigma -2k_1^{\alpha_2}
\delta^{\alpha_1}_{\sigma}\big)\cr
&\hskip100pt\cdot
\big(\eta^{\alpha_3\alpha_4}(k_3-k_4)^\sigma
+ 2k_4^{\alpha_3}\eta^{\sigma\alpha_4}
-2 k_3^{\alpha_4}\eta^{\sigma\alpha_3}\big)\cr
&\;  + {1\over s} \;\sum_{1\;\le i<j\le 4}\Big( P_i^\mu D_j
+ P_{i\rho}L_j^{\mu\rho}- (i\leftrightarrow j)
\Big)^{\alpha_1\alpha_2\alpha_3\alpha_4}_{\;\gamma_1\gamma_2\gamma_3\gamma_4}
\cr &\hskip40pt
\cdot\big(\eta^{\gamma_1\gamma_2}(k_1-k_2)_\sigma
+ 2 k_2^{\gamma_1}\delta^{\gamma_2}_\sigma -2k_1^{\gamma_2}
\delta^{\gamma_1}_{\sigma}\big)
\big(\eta^{\gamma_3\gamma_4}(k_3-k_4)^\sigma
+ 2k_4^{\gamma_3}\eta^{\sigma\gamma_4}
-2 k_3^{\gamma_4}\eta^{\sigma\gamma_3}\big)\cr
\label{PAAAA}\end{align}
\begin{align}
&-\hat P^\mu \; i {n_t\over t} \cr&=
\sum_{1\;\le i<j\le 4}\Big( P_i^\mu D_j + P_{i\rho}L_j^{\mu\rho}
- (i\leftrightarrow j)
\Big)^{\alpha_1\alpha_2\alpha_3\alpha_4}_{\;\gamma_1\gamma_2\gamma_3\gamma_4}
\;\;  {i\over t} \; n_t^{\gamma_1\gamma_2\gamma_3\gamma_4}\cr
&= \sum_{1\;\le i<j\le 4}\Big( P_i^\mu D_j + P_{i\rho}L_j^{\mu\rho}
- (i\leftrightarrow j)
\Big)^{\alpha_1\alpha_2\alpha_3\alpha_4}_{\;\gamma_1\gamma_2\gamma_3\gamma_4}
\;\;(-\eta^{\gamma_1\gamma_3}\eta^{\gamma_2\gamma_4}
+\eta^{\gamma_1\gamma_2}\eta^{\gamma_3\gamma_4})\cr
&+ 2(-k_1+k_4)^\mu\;\;  {1\over t} \;
\big(\eta^{\alpha_2\alpha_3}(k_2-k_3)_\sigma
+ 2 k_3^{\alpha_2}\delta^{\alpha_3}_\sigma -2k_2^{\alpha_3}
\delta^{\alpha_2}_{\sigma}\big)\cr&\hskip100pt\cdot
\big(\eta^{\alpha_4\alpha_1}(k_1-k_4)^\sigma
- 2k_1^{\alpha_4}\eta^{\sigma\alpha_1}
+ 2 k_4^{\alpha_1}\eta^{\sigma\alpha_4}\big)\cr
&\;  + {1\over t} \;\sum_{1\;\le i<j\le 4}\Big( P_i^\mu D_j
+ P_{i\rho}L_j^{\mu\rho}- (i\leftrightarrow j)
\Big)^{\alpha_1\alpha_2\alpha_3\alpha_4}_{\;\gamma_1\gamma_2\gamma_3\gamma_4}
\cr &\hskip50pt
\cdot\big(\eta^{\gamma_2\gamma_3}(k_2-k_3)_\sigma
+ 2 k_3^{\gamma_2}\delta^{\gamma_3}_\sigma -2k_2^{\gamma_3}
\delta^{\gamma_2}_{\sigma}\big)
\big(\eta^{\gamma_4\gamma_1}(k_1-k_4)^\sigma
- 2k_1^{\gamma_4}\eta^{\sigma\gamma_1}
+ 2 k_4^{\gamma_1}\eta^{\sigma\gamma_4}\big)\cr
\label{PBBBB}\end{align}
Evaluating (\ref{PAAAA}), (\ref{PBBBB}) using momentum conservation, dropping
terms that are gauge transformations {\it i.e.} proportional to
$k_i^{\alpha_i}$,
and using the on-shell conditions $k_i^2=0$, yields (\ref{summary1}).

\section{\hskip10pt \bf Supercharge Contribution for 
the Four-Point Gauge Amplitude}
\label{SCG4}
For the four-point function we motivate the identity between
(\ref{summary1}) and (\ref{QQQQ})
by adding supercharges to the level one generator for $N=4$,
\begin{align}
&\langle 0|T A^{\gamma_1}(x_1) A^{\gamma_2}(x_2) A^{\gamma_3}(x_3)
A^{\gamma_4}(x_4) |0\rangle =
G^{\gamma_1\gamma_2\gamma_3\gamma_4}(x_1 x_2 x_3 x_4)\cr
&\hbox{For super Yangian invariance,}\cr
&-\hat P^{\mu\alpha_1\alpha_2\alpha_3\alpha_4}_{x,SS\;\gamma_1\gamma_2
\gamma_3\gamma_4}
\; G^{\gamma_1\gamma_2\gamma_3\gamma_4}(x_1 x_2 x_3 x_4) = 0
\cr&=\hskip-5pt
\sum_{1\;\le i<j\le 4}\Big( P_i^\mu D_j + P_{i\rho}L_j^{\mu\rho}
- (i\leftrightarrow j) \cr
&\hskip90pt
-\quar \bar\sigma^{\mu\dot\alpha\alpha}
Q^A_{\alpha i} \tilde Q_{A\dot\alpha j} - (i\leftrightarrow j)
\Big )^{\alpha_1\alpha_2\alpha_3\alpha_4}_{x\;\gamma_1\gamma_2\gamma_3\gamma_4}
G^{\gamma_1\gamma_2\gamma_3\gamma_4}(x_1 x_2 x_3 x_4)\cr
&=\sum_{1\;\le i<j\le 4}\Big( P_i^\mu D_j + P_{i\rho}L_j^{\mu\rho}
- (i\leftrightarrow j)
\Big)^{\alpha_1\alpha_2\alpha_3\alpha_4}_{x\;\gamma_1\gamma_2\gamma_3\gamma_4}
\;  G^{\gamma_1\gamma_2\gamma_3\gamma_4}(x_1 x_2 x_3 x_4)\cr
&-\quar\bar\sigma^{\mu\dot\alpha\alpha}
\langle 0| T Q^A_{\alpha 1}  A^{\alpha_1}(x_1)
\tilde Q_{A\dot\alpha 2}  A^{\alpha_2}(x_2) A^{\alpha_3}(x_3)
A^{\alpha_4}(x_4)|0\rangle
\cr&-\quar\bar\sigma^{\mu\dot\alpha\alpha}
\langle 0| T \tilde Q_{A\dot\alpha 1}  A^{\alpha_1}(x_1)
Q^A_{\alpha 2}  A^{\alpha_2}(x_2) A^{\alpha_3}(x_3) A^{\alpha_4}(x_4)
|0\rangle\cr
&-\quar\bar\sigma^{\mu\dot\alpha\alpha}
\langle 0| T Q^A_{\alpha 1}  A^{\alpha_1}(x_1)
A^{\alpha_2}(x_2) \tilde Q_{A\dot\alpha 3} A^{\alpha_3}(x_3) A^{\alpha_4}(x_4)
|0\rangle\cr
&-\quar\bar\sigma^{\mu\dot\alpha\alpha}
\langle 0| T \tilde Q_{A\dot\alpha 1}  A^{\alpha_1}(x_1)
A^{\alpha_2}(x_2) Q^A_{\alpha 3}  A^{\alpha_3}(x_3) A^{\alpha_4}(x_4)
|0\rangle\cr
&-\quar\bar\sigma^{\mu\dot\alpha\alpha}
\langle 0| T A^{\alpha_1}(x_1)
Q^A_{\alpha 2} A^{\alpha_2}(x_2)
\tilde Q_{A\dot\alpha 3} A^{\alpha_3}(x_3) A^{\alpha_4}(x_4)|0\rangle\cr
&-\quar\bar\sigma^{\mu\dot\alpha\alpha}
\langle 0| T A^{\alpha_1}(x_1)
\tilde Q_{A\dot\alpha 2} A^{\alpha_2}(x_2) Q^A_{\alpha 3}
A^{\alpha_3}(x_3) A^{\alpha_4}(x_4)|0\rangle
\cr
\cr
&-\quar\bar\sigma^{\mu\dot\alpha\alpha}
\langle 0| T Q^A_{\alpha 1}  A^{\alpha_1}(x_1)
A^{\alpha_2}(x_2) A^{\alpha_3}(x_3)
\tilde Q_{A\dot\alpha 4}A^{\alpha_4}(x_4)|0\rangle
\cr
&-\quar\bar\sigma^{\mu\dot\alpha\alpha}
\langle 0| T \tilde Q_{A\dot\alpha 1}  A^{\alpha_1}(x_1)
A^{\alpha_2}(x_2) A^{\alpha_3}(x_3) Q^A_{\alpha 4}A^{\alpha_4}(x_4)
|0\rangle\cr
&-\quar\bar\sigma^{\mu\dot\alpha\alpha}
\langle 0| T A^{\alpha_1}(x_1)
Q^A_{\alpha 2} A^{\alpha_2}(x_2)
A^{\alpha_3}(x_3) \tilde Q_{A\dot\alpha 4}A^{\alpha_4}(x_4)|0\rangle\cr
&-\quar\bar\sigma^{\mu\dot\alpha\alpha}
\langle 0| T A^{\alpha_1}(x_1)
\tilde Q_{A\dot\alpha 2} A^{\alpha_2}(x_2)
A^{\alpha_3}(x_3)  Q^A_{\alpha 4} A^{\alpha_4}(x_4)|0\rangle\cr
&-\quar\bar\sigma^{\mu\dot\alpha\alpha}
\langle 0| T A^{\alpha_1}(x_1)
A^{\alpha_2}(x_2) Q^A_{\alpha 3}  A^{\alpha_3}(x_3) \tilde Q_{A\dot\alpha 4}
A^{\alpha_4}(x_4)
|0\rangle\cr
&-\quar\bar\sigma^{\mu\dot\alpha\alpha}
\langle 0| T A^{\alpha_1}(x_1)
A^{\alpha_2}(x_2) \tilde Q_{A\dot\alpha 3} A^{\alpha_3}(x_3)
 Q^A_{\alpha 4} A^{\alpha_4}(x_4)
|0\rangle\cr
\label{4TQQAAA}\end{align}
Then working to second order in the coupling $g$,
\begin{align}
&-\quar\bar\sigma^{\mu\dot\alpha\alpha}
\langle 0| T Q^A_{\alpha 1}  A^{a\alpha_1}(x_1)
\tilde Q_{A\dot\alpha 2}  A^{b\alpha_2}(x_2)
A^{c\alpha_3}(x_3)  A^{d\alpha_4}(x_4)|0\rangle
\cr &= -\quar\bar\sigma^{\mu\dot\alpha\alpha}\;
\sigma^{\alpha_1}_{\alpha\dot\beta} \epsilon^{\dot\beta\dot\gamma}
\epsilon_{\dot\alpha\dot\kappa}\bar
\sigma^{\alpha_2\dot\kappa\gamma}
\langle 0| T \bar\psi^{Aa}_{\dot\gamma}(x_1)
\psi^b_{A\gamma}(x_2)A^{c\alpha_3}(x_3)
A^{d\alpha_4}(x_4) e^{i\int d^4 z {\cal L}_I}|0\rangle
\label{twotime}\\
&=  - g^2{\textstyle{1\over 8}}\bar\sigma^{\mu\dot\alpha\alpha}\;
\sigma^{\alpha_1}_{\alpha\dot\beta} \epsilon^{\dot\beta\dot\gamma}
\epsilon_{\dot\alpha\dot\kappa}\bar
\sigma^{\alpha_2\dot\kappa\gamma} \int d^4 z_1 d^4 z_2
\langle 0 |T
\bar\psi^{Aa}_{\dot\gamma}(x_1)
\psi^{b}_{A\gamma}(x_2)A^{c\alpha_3}(x_3)
A^{d\alpha_4}(x_4)\cr 
&\hskip90pt\cdot
\bar\psi^{Bm}_{\dot\delta}(z_1)
\bar\sigma^{\nu\dot\delta\delta}A^{e}_\nu (z_1)
\psi^f_{B\delta}(z_1)
\;\bar\psi^{Ch}_{\dot\epsilon}(z_2)
\bar\sigma^{\rho\dot\epsilon\epsilon}A^j_\rho (z_2)
\psi^{\ell}_{C\epsilon}(z_2)
|0\rangle\;f_{mef} f_{hj\ell}\cr
&= \half g^2  \bar\sigma^{\mu\dot\alpha\alpha}\;
\sigma^{\alpha_1}_{\alpha\dot\beta} \epsilon^{\dot\beta\dot\gamma}
\epsilon_{\dot\alpha\dot\kappa}\bar
\sigma^{\alpha_2\dot\kappa\gamma} \bar\sigma^{\nu\dot\delta\delta}
\bar\sigma^{\rho\dot\epsilon\epsilon}
\int d^4 z_1 d^4 z_2 \cr
&\hskip7pt\cdot \big[
D^{\alpha_3}_\nu(x_3-z_1) D^{\alpha_4}_\rho(x_4-z_2)
S^F_{\epsilon\dot\gamma}(z_2-x_1) S^F_{\gamma\dot\delta}(x_2-z_1)
S^F_{\delta\dot\epsilon}(z_1-z_2)\;f_{bch} f_{hda}\cr
&\hskip15pt
+ D^{\alpha_3}_\nu(x_3-z_1) D^{\alpha_4}_\rho(x_4-z_2)
S^F_{\delta\dot\gamma}(z_1-x_1)
S^F_{\gamma\dot\epsilon}(x_2-z_2)
S^F_{\epsilon\dot\delta}(z_2-z_1)\;
f_{\ell ca} f_{bd\ell}\cr
&\hskip15pt
+  D^{\alpha_3}_\rho(x_3-z_2) D^{\alpha_4}_\nu(x_4-z_1)
S^F_{\epsilon\dot\gamma}(z_2-x_1)
S^F_{\gamma\dot\delta}(x_2-z_1)
S^F_{\delta\dot\epsilon}(z_1-z_2)\;
f_{bdh} f_{hca}\cr
&\hskip15pt
+  D^{\alpha_3}_\rho(x_3-z_2) D^{\alpha_4}_\nu(x_4-z_1)
S^F_{\delta\dot\gamma}(z_1-x_1)
S^F_{\gamma\dot\epsilon}(x_2-z_2)
S^F_{\epsilon\dot\delta}(z_2-z_1) \;
f_{\ell da} f_{bc\ell}\big]
\label{4Firstterm}\end{align}
with ${\cal L}_I$ given in (\ref{Linter}). As in section \ref{hatPG3},
in this Appendix
$1\le \alpha,\dot\alpha\le 2$ and $0\le \alpha_i\le 3$ for site $i$.
With multiplication by four inverse gluon propagators,
the truncated Fourier transform of (\ref{4Firstterm}) is
\begin{align}
& -i g^2 (2\pi)^4\delta^4(k_1+k_2+k_3+k_4)\cr&\hskip7pt \cdot\Big(
(-\bar\sigma^{\tilde\alpha_1}\sigma^\mu\bar\sigma^{\tilde\alpha_2}
\sigma^\zeta\bar\sigma^{\tilde\alpha_3}\sigma^\tau\bar\sigma^{\tilde
\alpha_4}\sigma^\omega)^{\dot\gamma}_{\hskip 3pt\dot\gamma}\;
k_{1\omega}k_{2\zeta} (k_2+k_3)_\tau \;{1\over (k_2+k_3)^2}\;
f_{bce} f_{eda}\cr
&\hskip20pt 
+(-\bar\sigma^{\tilde\alpha_1}\sigma^\mu\bar\sigma^{\tilde\alpha_2}
\sigma^\zeta\bar\sigma^{\tilde\alpha_4}\sigma^\tau\bar\sigma^{\tilde
\alpha_3}\sigma^\omega)^{\dot\gamma}_{\hskip 3pt\dot\gamma}\;
k_{1\omega}k_{2\zeta} (k_2+k_4)_\tau \;{1\over (k_2+k_4)^2}\;
\;f_{cae} f_{ebd}\Big)
\label{ft1}\end{align}
There is also a contribution to (\ref{twotime})
from the interaction Lagrangian
given by
\begin{align}
&-\quar\bar\sigma^{\mu\dot\alpha\alpha}
\langle 0| T Q^A_{\alpha 1}  A^{a\alpha_1}(x_1)
\tilde Q_{A\dot\alpha 2}  A^{b\alpha_2}(x_2)
A^{ c\alpha_3}(x_3)  A^{d\alpha_4}(x_4)|0\rangle
\cr &= -\quar\bar\sigma^{\mu\dot\alpha\alpha}\;
\sigma^{\alpha_1}_{\alpha\dot\beta} \epsilon^{\dot\beta\dot\gamma}
\epsilon_{\dot\alpha\dot\kappa}\bar
\sigma^{\alpha_2\dot\kappa\gamma}
\langle 0| T \bar\psi^{Aa}_{\dot\gamma}(x_1)
\psi^{b}_{A\gamma}(x_2)A^{c\alpha_3}(x_3)
A^{d\alpha_4}(x_4) e^{i\int d^4 z {\cal L}_I}|0\rangle\cr
&=  i g^2\quar \bar\sigma^{\mu\dot\alpha\alpha}\;
\sigma^{\alpha_1}_{\alpha\dot\beta} \epsilon^{\dot\beta\dot\gamma}
\epsilon_{\dot\alpha\dot\kappa}\bar
\sigma^{\alpha_2\dot\kappa\gamma} \int d^4 z_1 d^4 z_2
\langle 0 |T
\bar\psi^{Aa}_{\dot\gamma}(x_1)
\psi^{b}_{A\gamma}(x_2)A^{c\alpha_3}(x_3)
A^{d\alpha_3}(x_4)\cr &\hskip90pt \cdot
\bar\psi^{Bm}_{\dot\delta}(z_1)
\bar\sigma^{\nu\dot\delta\delta}A^e_\nu (z_1)
\psi^{f}_{B\delta}(z_1)
\;A_\rho^j(z_2) A_\sigma^\ell(z_2) \partial^\rho A^{\sigma h}(z_2)
|0\rangle\;f_{mef} f_{hj\ell}\cr
&=i g^2  \bar\sigma^{\mu\dot\alpha\alpha}\;
\sigma^{\alpha_1}_{\alpha\dot\beta} \epsilon^{\dot\beta\dot\gamma}
\epsilon_{\dot\alpha\dot\kappa}\bar
\sigma^{\alpha_2\dot\kappa\gamma} \bar\sigma^{\nu\dot\delta\delta}
\int d^4 z_1 d^4 z_2 \;
 S^F_{\delta\dot\gamma} (z_1-x_1)
S^F_{\gamma\dot\delta} (x_2-z_1) \;f_{bea}\cr
&\hskip7pt \cdot \big[D^{\alpha_3}_\rho(x_3-z_2)
D^{\alpha_4}_\sigma(x_4-z_2)\partial^\rho_{z_2} D^\sigma_\nu(z_1-z_2)
\cr&\hskip20pt
-D^{\alpha_3}_\rho(x_3-z_2)
\partial^\rho_{z_2}
D^{\alpha_4\sigma} (x_4-z_2) D_{\nu\sigma} (z_1-z_2)
\cr
&\hskip20pt
-D^{\alpha_3}_\sigma (x_3-z_2)
D^{\alpha_4}_\rho (x_4-z_2) \partial^\rho_{z_2}D_\nu^\sigma (z_1-z_2)
\cr&\hskip20pt
+D^{\alpha_3}_\sigma (x_3-z_2)
\partial^\rho_{z_2} D^{\alpha_4\sigma}(x_4-z_2)
D_{\nu\rho}(z_1-z_2)
\cr&\hskip20pt
+\partial^\rho_{z_2} D^{\alpha_3\sigma} (x_3-z_2)
D^{\alpha_4}_\rho (x_4-z_2) D_{\nu\sigma}(z_1-z_2)
\cr&\hskip20pt
-\partial^\rho_{z_2} D^{\alpha_3\sigma} (x_3-z_2)
D^{\alpha_4}_\sigma (x_4-z_2)
D_{\nu\rho}(z_1-z_2) \Big] f_{ecd}\cr
\label{extrainter}
\end{align}
The truncated Fourier transform of (\ref{extrainter}) is
\begin{align}
& -i g^2 (2\pi)^4 \delta^4(k_1+k_2+k_3+k_4)\;f_{bea}f_{ecd}\cr
&\hskip20pt\cdot
\Big[ (\bar\sigma^{\tilde\alpha_1}\sigma^\mu\bar\sigma^{\tilde\alpha_2}
\sigma^\tau\bar\sigma^{\tilde\alpha_4}\sigma^\omega)^{\dot\gamma}_{\dot\gamma}
\; {k_{1\omega} k_{2\tau}\over (k_3+k_4)^2}\;
\Big(-k_3-2k_4 \Big)^{\tilde\alpha_3} \cr
&\hskip 30pt +
(\bar\sigma^{\tilde\alpha_1}\sigma^\mu\bar\sigma^{\tilde\alpha_2}
\sigma^\tau\bar\sigma^{\tilde\alpha_3}\sigma^\omega)^{\dot\gamma}_{\dot\gamma}
\; {k_{1\omega} k_{2\tau}\over (k_3+k_4)^2}\;
\Big(2k_3+k_4\Big)^{\tilde\alpha_4}
\cr&\hskip30pt
+ (\bar\sigma^{\tilde\alpha_1}\sigma^\mu\bar\sigma^{\tilde\alpha_2}
\sigma^\tau\bar\sigma^\nu\sigma^\omega)^{\dot\gamma}_{\dot\gamma}
\; {k_{1\omega} k_{2\tau}\;\over (k_3+k_4)^2}\;\eta^{\tilde\alpha_3\tilde
\alpha_4}
\Big(-k_3+k_4 \Big)_\nu\Big]\cr
\label{4sec}\end{align}
The second supercharge term in (\ref{4TQQAAA})
is evaluated in a similar way, which when added to (\ref{ft1}) and
(\ref{4sec}) will promote the sigma matrices to Dirac matrices
(\ref{gammatr2}). The truncated Fourier transform of
\begin{align}
&-\quar\bar\sigma^{\mu\dot\alpha\alpha}
\langle 0| T Q^A_{\alpha 1}  A^{a\alpha_1}(x_1)
\tilde Q_{A\dot\alpha 2}  A^{b\alpha_2}(x_2)
A^{ c\alpha_3}(x_3)  A^{d\alpha_4}(x_4)|0\rangle
\cr
&-\quar\bar\sigma^{\mu\dot\alpha\alpha}
\langle 0| T \tilde Q_{A\dot\alpha 1}  A^{a \alpha_1}(x_1)
Q^A_{\alpha 2}  A^{b\alpha_2}(x_2)
A^{c\alpha_3}(x_3) A^{d\alpha_4}(x_4)|0\rangle
\label{qaqa}\end{align}
is
\begin{align}
&ig^2  (2\pi)^4\delta^4(k_1+k_2+k_3+k_4)\;
tr (\gamma^{\tilde\alpha_1}\gamma^\mu
\gamma^{\tilde\alpha_2}
\gamma^\zeta\gamma^{\tilde\alpha_3}\gamma^\tau\gamma^{\tilde
\alpha_4}\gamma^\omega)
\;k_{1\omega}k_{2\zeta} {(k_2+k_3)_\tau \over (k_2+k_3)^2}\;
f_{bce} f_{eda}\cr
&ig^2  (2\pi)^4\delta^4(k_1+k_2+k_3+k_4)\;
tr (\gamma^{\tilde\alpha_1}\gamma^\mu
\gamma^{\tilde\alpha_2}
\gamma^\zeta\gamma^{\tilde\alpha_4}\gamma^\tau\gamma^{\tilde
\alpha_3}\gamma^\omega)
\;k_{1\omega}k_{2\zeta} {(k_2+k_4)_\tau \over (k_2+k_4)^2}\;
f_{cae} f_{ebd}\cr
&-i g^2 (2\pi)^4 \delta^4(k_1+k_2+k_3+k_4)\;f_{bea}f_{ecd}
\; \Bigg[ tr (\gamma^{\tilde\alpha_1}\gamma^\mu\gamma^{\tilde\alpha_2}
\gamma^\tau\gamma^{\tilde\alpha_4}\gamma^\omega)
\; {k_{1\omega} k_{2\tau}\over (k_3+k_4)^2}\;
\Big(-2k_4^{\tilde\alpha_3}\Big) \cr
&\hskip 100pt +
tr (\gamma^{\tilde\alpha_1}\gamma^\mu\gamma^{\tilde\alpha_2}
\gamma^\tau\gamma^{\tilde\alpha_3}\gamma^\omega)
\; {k_{1\omega} k_{2\tau}\over (k_3+k_4)^2}\;
\Big(2k_3^{\tilde\alpha_4}\Big)
\cr &\hskip100pt
+ tr (\gamma^{\tilde\alpha_1}\gamma^\mu\gamma^{\tilde\alpha_2}
\gamma^\tau\gamma^\nu\gamma^\omega)
\; {k_{1\omega} k_{2\tau}\;\over (k_3+k_4)^2}\;\eta^{\tilde\alpha_3\tilde
\alpha_4}
\Big(-k_3+k_4 \Big)_\nu\Bigg]\cr
\label{ftos}\end{align}
where also we could have used the antisymmetry under $a\leftrightarrow b$,
$x_1\leftrightarrow x_2$, $\tilde\alpha_1\leftrightarrow \tilde\alpha_2$,
to generate the second term from the first term in (\ref{qaqa}).
We have dropped gauge transformations, {\it i.e.}
those terms proportional to $k_i^{\tilde\alpha_i}$.

Analogous symmetries can be used to generate the remaining ten supercharge
terms in (\ref{4TQQAAA}) from the first two (\ref{ftos}),
to find that total supercharge contribution to  (\ref{4TQQAAA}) is
\vfill\eject
\thispagestyle{empty}
\begin{align}
& i g^2 \; (2\pi)^4\delta^4(k_1+k_2+k_3+k_4)\cr
&\cdot\Bigg( f_{bce} f_{ead}\;
{(k_2+k_3)_\tau\over (k_2+k_3)^2}\;
\Big[ tr (\gamma^{\tilde\alpha_1}\gamma^\mu\gamma^{\tilde\alpha_2}
\gamma^\zeta\gamma^{\tilde\alpha_3}\gamma^\tau\gamma^{\tilde
\alpha_4}\gamma^\omega) k_{1\omega}k_{2\zeta}
- tr (\gamma^{\tilde\alpha_1}\gamma^\mu\gamma^{\tilde\alpha_3}
\gamma^\zeta\gamma^{\tilde\alpha_2}\gamma^\tau\gamma^{\tilde
\alpha_4}\gamma^\omega) k_{1\omega}k_{3\zeta} \cr
&\hskip 80pt + tr (\gamma^{\tilde\alpha_2}\gamma^\mu\gamma^{\tilde\alpha_4}
\gamma^\zeta\gamma^{\tilde\alpha_1}\gamma^\tau\gamma^{\tilde
\alpha_3}\gamma^\omega) k_{2\omega}k_{4\zeta}
- tr (\gamma^{\tilde\alpha_3}\gamma^\mu\gamma^{\tilde\alpha_4}
\gamma^\zeta\gamma^{\tilde\alpha_1}\gamma^\tau\gamma^{\tilde
\alpha_2}\gamma^\omega) k_{3\omega}k_{4\zeta}\Big] \cr
&+ f_{abe} f_{ecd}\;
{(k_1+k_2)_\tau\over (k_1+k_2)^2}\;
\Big[ -tr (\gamma^{\tilde\alpha_1}\gamma^\mu\gamma^{\tilde\alpha_3}
\gamma^\zeta\gamma^{\tilde\alpha_4}\gamma^\tau\gamma^{\tilde
\alpha_2}\gamma^\omega) k_{1\omega}k_{3\zeta}
+ tr (\gamma^{\tilde\alpha_2}\gamma^\mu\gamma^{\tilde\alpha_3}
\gamma^\zeta\gamma^{\tilde\alpha_4}\gamma^\tau\gamma^{\tilde
\alpha_1}\gamma^\omega) k_{2\omega}k_{3\zeta}\cr
&\hskip 80pt + tr (\gamma^{\tilde\alpha_1}\gamma^\mu\gamma^{\tilde\alpha_4}
\gamma^\zeta\gamma^{\tilde\alpha_3}\gamma^\tau\gamma^{\tilde
\alpha_2}\gamma^\omega) k_{1\omega}k_{4\zeta}
- tr (\gamma^{\tilde\alpha_2}\gamma^\mu\gamma^{\tilde\alpha_4}
\gamma^\zeta\gamma^{\tilde\alpha_3}\gamma^\tau\gamma^{\tilde
\alpha_1}\gamma^\omega) k_{2\omega}k_{4\zeta}\Big]\cr
&+f_{cae} f_{ebd}\;
{(k_1+k_3)_\tau\over (k_1+k_3)^2}\;
\Big[ tr (\gamma^{\tilde\alpha_1}\gamma^\mu\gamma^{\tilde\alpha_2}
\gamma^\zeta\gamma^{\tilde\alpha_4}\gamma^\tau\gamma^{\tilde
\alpha_3}\gamma^\omega) k_{1\omega}k_{2\zeta}
+ tr (\gamma^{\tilde\alpha_2}\gamma^\mu\gamma^{\tilde\alpha_3}
\gamma^\zeta\gamma^{\tilde\alpha_1}\gamma^\tau\gamma^{\tilde
\alpha_4}\gamma^\omega) k_{2\omega}k_{3\zeta}\cr
&\hskip 80pt - tr (\gamma^{\tilde\alpha_1}\gamma^\mu\gamma^{\tilde\alpha_4}
\gamma^\zeta\gamma^{\tilde\alpha_2}\gamma^\tau\gamma^{\tilde
\alpha_3}\gamma^\omega) k_{1\omega}k_{4\zeta}
+ tr (\gamma^{\tilde\alpha_3}\gamma^\mu\gamma^{\tilde\alpha_4}
\gamma^\zeta\gamma^{\tilde\alpha_2}\gamma^\tau\gamma^{\tilde
\alpha_1}\gamma^\omega) k_{3\omega}k_{4\zeta}\Big]\cr
& -\;
f_{abe}f_{ecd} \;
\Bigg[{k_{1\omega} k_{2\tau}\over (k_3+k_4)^2}
\; \Big[ -2 k_4^{\tilde\alpha_3}\;
tr (\gamma^{\tilde\alpha_1}\gamma^\mu\gamma^{\tilde\alpha_2}
\gamma^\tau\gamma^{\tilde\alpha_4}\gamma^\omega) + 2k_3^{\tilde\alpha_4}
tr (\gamma^{\tilde\alpha_1}\gamma^\mu\gamma^{\tilde\alpha_2}
\gamma^\tau\gamma^{\tilde\alpha_3}\gamma^\omega)
\cr&\hskip200pt +  \eta^{\tilde\alpha_3\tilde\alpha_4}
\Big(-k_3+k_4 \Big)_\nu\;
tr (\gamma^{\tilde\alpha_1}\gamma^\mu\gamma^{\tilde\alpha_2}
\gamma^\tau\gamma^\nu\gamma^\omega)\Big]\cr
&\hskip80pt  + {k_{3\omega} k_{4\tau}\over (k_3+k_4)^2}
\Big[-2 k_2^{\tilde\alpha_1}\;
tr (\gamma^{\tilde\alpha_3}\gamma^\mu\gamma^{\tilde\alpha_4}
\gamma^\tau\gamma^{\tilde\alpha_2}\gamma^\omega)
+ 2k_1^{\tilde\alpha_2}
tr (\gamma^{\tilde\alpha_3}\gamma^\mu\gamma^{\tilde\alpha_4}
\gamma^\tau\gamma^{\tilde\alpha_1}\gamma^\omega)
\cr&\hskip 200pt +  \eta^{\tilde\alpha_1\tilde\alpha_2}
\Big(-k_1+k_2 \Big)_\nu\;
tr (\gamma^{\tilde\alpha_3}\gamma^\mu\gamma^{\tilde\alpha_4}
\gamma^\tau\gamma^\nu\gamma^\omega)
\Big]\Bigg]\cr
&
-\;f_{bce}f_{ead} \;
\Bigg[{k_{2\omega} k_{3\tau}\over (k_1+k_4)^2}
\; \Big[ -2 k_4^{\tilde\alpha_1}\;
tr (\gamma^{\tilde\alpha_2}\gamma^\mu\gamma^{\tilde\alpha_3}
\gamma^\tau\gamma^{\tilde\alpha_4}\gamma^\omega)
+ 2k_1^{\tilde\alpha_4}
\;tr (\gamma^{\tilde\alpha_2}\gamma^\mu\gamma^{\tilde\alpha_3}
\gamma^\tau\gamma^{\tilde\alpha_1}\gamma^\omega)
\cr&\hskip200pt +  \eta^{\tilde\alpha_1\tilde\alpha_4}
\Big(-k_1+k_4 \Big)_\nu\;
tr (\gamma^{\tilde\alpha_2}\gamma^\mu\gamma^{\tilde\alpha_3}
\gamma^\tau\gamma^\nu\gamma^\omega)\Big]\cr
&\hskip90pt + {k_{1\omega} k_{4\tau}\over (k_2+k_3)^2}
\Big[-2 k_3^{\tilde\alpha_2}\;
tr (\gamma^{\tilde\alpha_1}\gamma^\mu\gamma^{\tilde\alpha_4}
\gamma^\tau\gamma^{\tilde\alpha_3}\gamma^\omega)
+ 2k_2^{\tilde\alpha_3}
tr (\gamma^{\tilde\alpha_1}\gamma^\mu\gamma^{\tilde\alpha_4}
\gamma^\tau\gamma^{\tilde\alpha_2}\gamma^\omega)
\cr&\hskip 200pt
+  \eta^{\tilde\alpha_2\tilde\alpha_3}
\Big(-k_2+k_3 \Big)_\nu\;
tr (\gamma^{\tilde\alpha_1}\gamma^\mu\gamma^{\tilde\alpha_4}
\gamma^\tau\gamma^\nu\gamma^\omega)
\Big]\Bigg]\cr
&-
\;f_{cae}f_{ebd} \;
\Bigg[{k_{1\omega} k_{3\tau}\over (k_2+k_4)^2}
\; \Big[ 2 k_4^{\tilde\alpha_2}\;
tr (\gamma^{\tilde\alpha_1}\gamma^\mu\gamma^{\tilde\alpha_3}
\gamma^\tau\gamma^{\tilde\alpha_4}\gamma^\omega)
- 2k_2^{\tilde\alpha_4}
\;tr (\gamma^{\tilde\alpha_1}\gamma^\mu\gamma^{\tilde\alpha_3}
\gamma^\tau\gamma^{\tilde\alpha_2}\gamma^\omega)
\cr&\hskip 200pt +  \eta^{\tilde\alpha_2\tilde\alpha_4}
\Big(k_2-k_4 \Big)_\nu\;
tr (\gamma^{\tilde\alpha_1}\gamma^\mu\gamma^{\tilde\alpha_3}
\gamma^\tau\gamma^\nu\gamma^\omega)\Big]\cr
&\hskip90pt + {k_{2\omega} k_{4\tau}\over (k_1+k_3)^2}
\Big[2 k_3^{\tilde\alpha_1}\;
tr (\gamma^{\tilde\alpha_2}\gamma^\mu\gamma^{\tilde\alpha_4}
\gamma^\tau\gamma^{\tilde\alpha_3}\gamma^\omega)
- 2k_1^{\tilde\alpha_3}
tr (\gamma^{\tilde\alpha_2}\gamma^\mu\gamma^{\tilde\alpha_4}
\gamma^\tau\gamma^{\tilde\alpha_1}\gamma^\omega)
\cr&\hskip 200pt+  \eta^{\tilde\alpha_1\tilde\alpha_3}
\Big(k_1-k_3 \Big)_\nu\;
tr (\gamma^{\tilde\alpha_2}\gamma^\mu\gamma^{\tilde\alpha_4}
\gamma^\tau\gamma^\nu\gamma^\omega)
\Big]\Bigg]\Bigg)\cr
&=  g^2  (2\pi)^4\delta^4(k_1+k_2+k_3+k_4)\;
\cdot\Big(f_{abe} f_{ecd}\;
\hat P^\mu\; {n_s\over s} +
f_{bce} f_{ead}\;
\hat P^\mu\; {n_t\over t}
+ f_{cae} f_{ebd}\;
\hat P^\mu\; {n_u\over u}\Big)
\cr\label{f112}\end{align}
Then the terms relating to $\hat P^\mu \; A(1234) \equiv
\hat P^\mu \; \Big( {i n_s\over s} - {i n_t\over t}\big) $
can be read off from (\ref{f112}), by taking the coefficient
of $f_{abe} f_{ecd}$ and subtracting from it the coefficient of
$f_{bce} f_{ead}$. That will result in $i$ times
the expression (\ref{QQQQ}) given in
section \ref{hatPG3}. This motivates the equivalence of
(\ref{summary1}) and (\ref{QQQQ}).

\section{Level One Yangian Generators in Component Form}
\label{Lev1}

Using (\ref{repi}) and the conformal
structure constants (\ref{structure}), we list 
expressions for all level one Yangian generators (\ref{muls}) in component form
\begin{align}
\hat L^{\mu\nu} &=  -{\half} \sum_{1\le i<j\le N}\big[\big(
P_i^\mu K_j^\nu - P_i^\nu K_j^\mu + 2 g_{\rho\sigma}
L_i^{\mu\rho} L_j^{\nu\sigma}\big)  - (i\leftrightarrow j)\big]\cr
\hat P^\mu &=  -
\sum_{1\le i<j\le N}\big[\big(P_i^\mu D_j
+ g_{\rho\sigma} P^\rho_i L^{\mu\sigma}_j\big) - (i\leftrightarrow j)\big]
\cr 
\hat K^\mu&= 
\sum_{1\le i<j\le N}\big[\big(-D_iK_j^\mu + g_{\rho\sigma}
L_i^{\mu\rho} K_j^\sigma\big) - (i\leftrightarrow j)\big]\cr
\hat D&= - {\half}\sum_{1\le i<j\le N}\big[g_{\rho\sigma} P_i^\rho
K_j^\sigma - (i\leftrightarrow j)\big]
\end{align}
For the momentum space differential operator representation,
$P_i^\mu, L_i^{\mu\nu}, K_i^\mu, D_i$ are given by (\ref{rep})
at site $i$.
\end{appendices}

\vskip20pt
\singlespacing

\vfil\eject

\providecommand{\bysame}{\leavevmode\hbox to3em{\hrulefill}\thinspace}
\providecommand{\MR}{\relax\ifhmode\unskip\space\fi MR }
\providecommand{\MRhref}[2]
{
}
\providecommand{\href}[2]{#2}

\end{document}